\documentclass{article}

\usepackage[final]{neurips_2024}

\usepackage{wrapfig}

\usepackage[utf8]{inputenc} 
\usepackage[T1]{fontenc}    
\usepackage{hyperref}       
\usepackage{url}            
\usepackage{booktabs}       
\usepackage{amsfonts}       
\usepackage{nicefrac}       
\usepackage{microtype}      
\usepackage{xcolor}         
\usepackage{graphicx}
\usepackage{algorithmicx}
\usepackage{algorithm}
\usepackage{amsmath}
\usepackage{enumitem}
\usepackage{pifont}
\usepackage{wrapfig}
\usepackage{diagbox}
\usepackage{enumitem}
\usepackage{multirow}

\usepackage{orcidlink}
\usepackage{algpseudocode}
\usepackage{amssymb}
\usepackage{bbm}
\usepackage{pifont}
\usepackage{hhline}
\usepackage{url}
\usepackage{todonotes}

\newtheorem{theorem}{Theorem}

\newcommand{\name}{\text{FedGMark}}

\title{{\name}: Certifiably Robust Watermarking for 
Federated Graph Learning} 

\author{%
Yuxin Yang$^{1,2}$\,\,\,\,Qiang Li$^{1}$\,\,\,\,Yuan Hong$^{3}$\,\,\,\,Binghui Wang$^{2}$\thanks{Corresponding Author (bwang70@iit.edu)}\\\\
$^1$College of Computer Science and
Technology, Jilin University, Changchun, Jilin, China \\
$^2$Department of Computer Science, Illinois Institute of Technology, Chicago, Illinois, USA \\
$^3$School of Computing, University of Connecticut, Storrs, Connecticut, USA\\
}

\begin{document}

\maketitle

\begin{abstract}
Federated graph learning (FedGL) is an emerging learning paradigm to collaboratively train   graph data from various clients. 
However, during the development and deployment of FedGL models,  they are  
susceptible to  
illegal copying and model theft. 
Backdoor-based watermarking is a {well-known} method for mitigating these attacks, as it offers ownership verification to the model owner. 
We take the first step to protect the ownership of FedGL models via backdoor-based watermarking. 
Existing 
techniques have challenges in achieving the goal:  
1) they either cannot be directly applied or yield unsatisfactory performance; 2) they are vulnerable to watermark removal attacks; and 
3) they lack of formal guarantees. 
To address all the challenges, we propose {\name}, the first certified robust backdoor-based watermarking for FedGL.  {\name} leverages the unique graph structure and client information in FedGL to learn customized and diverse watermarks. It also designs a novel GL architecture that facilitates defending against both the empirical and theoretically worst-case watermark removal attacks. 
Extensive experiments 
validate the promising empirical and provable watermarking performance of {\name}. 
{Source code is available at: \url{https://github.com/Yuxin104/FedGMark}. }
\end{abstract}
\section{Introduction}
\label{sec:introdution}
Federated Graph Learning (FedGL)~\citep{xie2021federated, wang2022graphfl, tan2023federated, yao2024fedgcn}  
leverages a server and multiple clients to collaboratively train GL methods~\citep{kipf2016semi, hamilton2017inductive}  
via federated learning (FL)~\citep{mcmahan2017communication, li2021ditto, karimireddy2020scaffold}. 
In recent years, 
FedGL has attracted increasing interest in domains such as disease prediction~\citep{peng2022fedni}, recommendation systems~\citep{baek2023personalized, wu2022federated, li2022federated}, and molecular classification~\citep{he2022spreadgnn}. In addition, several industries have deployed/open-sourced their FedGL frameworks, such as Alibaba's FederatedScope-GNN~\citep{wang2022federatedscope} and Amazon's FedML-GNN~\citep{Amazon_FedML_GNN}. 
However, FedGL models are typically left unprotected, rendering them vulnerable to threats like illegal copying, model theft, and 
malicious distribution. 
For instance, a business competitor may replicate a model to gain competitive advantages or a malicious user may sell the model for profits. 
These threats waste the model owner's investment (e.g., labor costs, time, and energy) and infringe upon the legitimate copyrights of the model. 

Backdoor-based watermarking~\citep{uchida2017embedding,adi2018turning, bansal2022certified} is a \emph{de facto} model ownership verification technique to mitigate the above threats. 
This technique typically consists of two steps: 1) Embedding the target model with a watermark. The model owner injects a specific backdoor trigger (i.e., watermark) into some clean samples and trains the target model with this watermarked data along with the remaining clean data. Then the trained target (watermarked) model could have both high watermark accuracy (i.e., accurately classify testing data with the same watermark as the owner desires) and main task accuracy (i.e., accurately classify clean testing data). 2) Model ownership verification. 
When suspecting the target model is illegally used by others,  
the model owner can recruit a trusted third party for model ownership verification. Particularly, the true model owner knows 
how the target model behaves as expected by providing the trusted third party the carefully designed watermarked data, while 
the illegal parties cannot do so. {Notice that, since all the clients have devoted computation and data to the training, they have a strong intention to jointly protect their ownership of the model.}

\begin{table}[!t]
\scriptsize
\renewcommand{\arraystretch}{1.1}
\addtolength{\tabcolsep}{-3.3pt}
\caption{Results of adapting the random graph-based watermarking GL method~\citep{xu2023watermarking} to watermark FedGL models.  
``MA'': main task accuracy; ``WA'': watermark accuracy.
}
\centering
    \begin{tabular}{c||c|c c c c||c|c c c}
    \toprule
    {Datasets}  & \multicolumn{1}{c}{Attack}&  {Fed-GIN}  &  {Fed-GSAGE} & {Fed-GCN} &{Datasets}  & \multicolumn{1}{c}{Attack}&  {Fed-GIN}  &  {Fed-GSAGE} & {Fed-GCN}  \\ 
    & & MA$\uparrow$ \ WA$\uparrow$ & MA$\uparrow$ \ WA$\uparrow$  & MA$\uparrow$ \ WA$\uparrow$  & & & MA$\uparrow$ \ WA$\uparrow$  & MA$\uparrow$ \ WA$\uparrow$  & MA$\uparrow$ \ WA$\uparrow$ \\
    \hline
    \hline
    & None & {\bf 0.82 \ \ 0.47} & {\bf 0.82 \ \ 0.42 } & {\bf 0.80 \ \ 0.43 } & &None & {\bf 0.71 \ \ 0.56  } & {\bf 0.70 \ \ 0.54} & {\bf 0.70 \ \ 0.52 } \\
    MUTAG&  Distillation & 0.81 \ \ 0.38  & 0.80 \ \ 0.35 &  0.75 \ \ 0.32 & PROTEINS &Distillation & 0.71 \ \ 0.30  & 0.70 \ \ 0.32 & 0.70 \ \ 0.28 \\
     & Finetuning& 0.82 \ \ 0.33  & 0.80 \ \ 0.29 & 0.78 \ \ 0.27  &  & Finetuning & 0.71 \ \ 0.27 & 0.70 \ \ 0.29  & 0.70 \ \ 0.25 \\
    & $1$-Layer Pert.  & 0.78 \ \ 0.24  & 0.79 \ \ 0.23  &  0.79 \ \ 0.19 & &$1$-Layer Pert.  &  0.67 \ \ 0.16 & 0.65 \ \ 0.15  & 0.68 \ \ 0.17   \\
    \hline
    & None & {\bf 0.73 \ \ 0.39} & {\bf 0.70 \ \ 0.38 } & {\bf 0.71 \ \ 0.38 } & &None & {\bf 0.73 \ \ 0.57 } & {\bf 0.71 \ \ 0.57 } & {\bf 0.72 \ \ 0.53} \\
    DD&  Distillation & 0.72 \ \ 0.32 &  0.70 \ \ 0.30 & 0.70 \ \ 0.30  & COLLAB &Distillation & 0.72 \ \ 0.47 &  0.73 \ \ 0.49 & 0.71 \ \ 0.47 \\
     & Finetuning&  0.72 \ \ 0.19 & 0.70 \ \ 0.21 &  0.70 \ \ 0.22 &  & Finetuning & 0.72 \ \ 0.35 &  0.72 \ \ 0.36 & 0.71 \ \ 0.40 \\
    & $1$-Layer Pert.  & 0.70 \ \ 0.11 & 0.70 \ \ 0.15  & 0.70 \ \ 0.12 & &$1$-Layer Pert.  & 0.67 \ \ 0.31  & 0.66 \ \ 0.26 & 0.68 \ \ 0.38  \\
    \bottomrule
    \end{tabular}
\vspace{-4mm}
\label{table:random-watermark}  
\end{table}

In this paper, we aim to protect the ownership of FedGL models via backdoor-based watermarking. 
We observe backdoor-based watermarking methods for protecting the ownership of FL model on \emph{non-graph} data~\citep{li2022fedipr, tekgul2021waffle,  shao2022fedtracker, yang2023watermarking, lansari2023federated} or centralized GL model on graph data~\citep{xu2023watermarking} have been recently developed.  
However, 
applying these methods for protecting FedGL models faces {challenges} and weaknesses.  
\begin{itemize}[leftmargin=*]
\vspace{-2mm}
\item \emph{Inapplicable or ineffective:} Existing methods for non-graph data cannot be directly applied for graph data. 
For instance, they require input data have same size, while graphs can have varying sizes; they are unable to consider the \emph{connectivity} information such as edges connecting nodes in the graph data. 
The only method for graph data~\citep{xu2023watermarking} uses a naive \emph{random graph} (e.g., generated by the ER model~\citep{gilbert1959random}) as a watermark.   
Extending this  
watermark from centralized GL  
to FedGL models exhibits \emph{unsatisfactory performance}, as shown in Table~\ref{table:random-watermark}. For instance, the watermark accuracy is less than 60\% in all the studied graph datasets and FedGL models. 
The core reason is the random graph watermark does not use any graph structure information or client information that are unique in FedGL.  

\item \emph{Vulnerable to watermark removal attacks:} They are vulnerable to existing watermark removal techniques such as distillation and finetuning~\citep{bansal2022certified} (more details in Section~\ref{sec:threatmodel}).  
For instance, as illustrated in Table~\ref{table:random-watermark}, 
distillation can reduce the watermark accuracy to less than 30\%. 

\item \emph{Lack or weak formal guarantees:} 
All these methods do not provide formal robustness guarantees against watermark removal attacks. This could make them even vulnerable to more advanced attacks. For instance, our proposed layer-perturbation attack can further reduce the watermark accuracy, e.g., perturbing only 1-layer parameters of the watermarked model yields only 10\% watermark accuracy (while main accuracy is marginally affected). 
\cite{bansal2022certified} proposed the first  certified watermark for centralized non-graph learning models against $l_2$ model parameter perturbation. 
However, its certified radius is only 1.2, meaning the $l_2$ norm of a (usually million-dimensional) perturbation vector cannot exceed 1.2 to maintain the watermark accuracy. 
\vspace{-2mm}
\end{itemize}

\begin{figure*}[!t]
\centering	
\includegraphics[width=\textwidth]{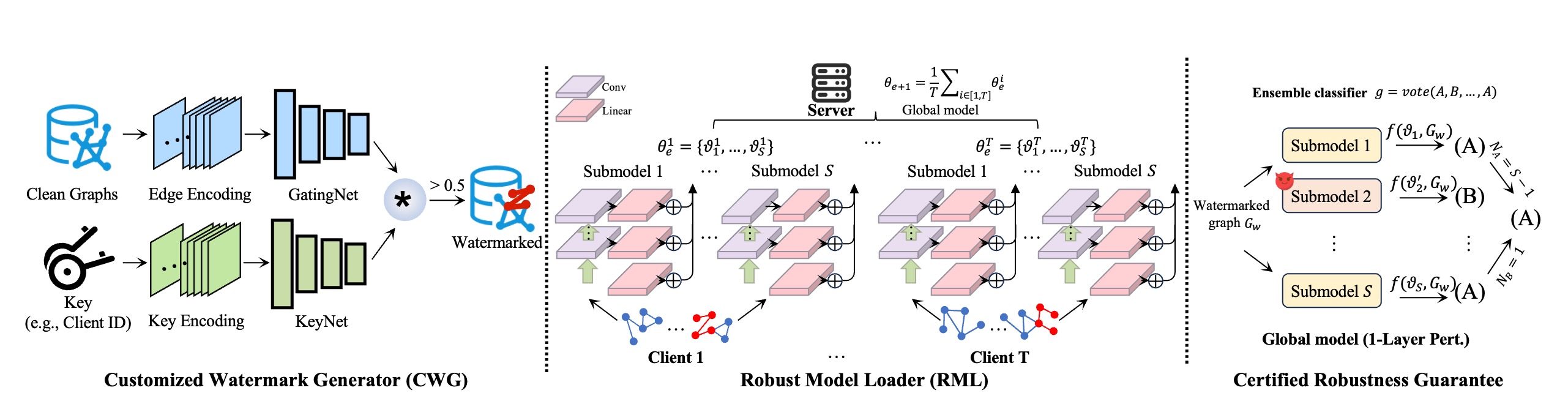}
    \vspace{-6mm}
	\caption{Overall pipeline of the proposed certified watermarks.} 
        \label{fig:overview}
    \vspace{-4mm}
\end{figure*}

We address all the above issues by proposing a certified robust backdoor-based watermark method for FedGL, called {\bf {\name}}.\footnote{In typical FL, a server and multiple clients collaboratively train a global model stored in the server, which is used by all clients for their tasks. Accordingly, in our ownership verification problem in FedGL, all clients design their own watermark data and collaboratively train the watermarked global model, which is for joint ownership by all participating clients. Note that we do not consider the case where the clients did not participate in watermark training, but claim the ownership of the model (actually these clients do not know how to do so).}  
{\name} 
enjoys several properties: 1) Its designed watermarks can handle varying size graphs and utilize both graph structure and client  information  
unique in FedGL models; 2) It is \emph{empirically} robust to both existing watermark removal attacks and the proposed layer-perturbation attack; and 3) more importantly, it is \emph{provably} robust to the  
layer-perturbation attack (the layer parameters can be arbitrarily perturbed), when the number of the perturbed layers  
is bounded. 
Specifically, as depicted in Figure~\ref{fig:overview}, {\name} consists of two modules: \emph{1) Customized Watermark Generator (CWG)}: it learns the customized watermark for individual graphs and clients in FedGL, by integrating the edge information from the client graphs and the  
unique key 
features of the clients. CWG can significantly enhance the diversity and effectiveness of the generated watermarks.  
\emph{2) Robust Model Loader (RML)}. 
RML designs a new  
GL model that consists of multiple submodels, where each submodel can be any existing GL model. 
It also
introduces a voting classifier for assembling the submodels’ predictions. Such a design can 
facilitate deriving the certified watermark performance against the (worst-case) layer-perturbation attack. 

We evaluate  
{\name} on four real-world graph datasets (MUTAG, PROTEINS, DD, and COLLAB) and three FedGL models including Fed-GIN, Fed-GSAGE, and Fed-GCN, whose base GL models are GIN~\citep{xu2018powerful}, GSAGE~\citep{hamilton2017inductive}, and GCN~\citep{kipf2016semi}, respectively.  
Extensive experimental results show {\name} achieves high main accuracy and watermark accuracy under no attacks and watermark removal attacks, high certified watermark accuracy, and significantly outperforms the existing method.  
Such good results demonstrate 
the potential of {\name}  as a watermarking method to protect the ownership of FedGL models. 

We summarize our main contributions of this paper as follows:
\begin{itemize}[leftmargin=*]
\vspace{-2mm}
\item To our best knowledge, this is the first work to protect the ownership of emerging FedGL models. 

\item We propose a certifiably robust backdoor-based watermarking method {\name} for FedGL.

\item We validate the effectiveness of {\name} in multiple  FedGL models and real-world graph datasets under no attack, existing backdoor removal attacks, and worst-case layer-perturbation attacks.  

\end{itemize}

\section{PRELIMINARIES}

\subsection{Federated Graph Learning (FedGL)}
\vspace{-2mm}
Given a graph $G = (\mathbb{V},\mathbb{E})$ as input, a GL model for graph classification learns a graph classifier $f$ that outputs a label $f(G)=y \in \mathbb{Y}$ for a graph. Here, $\mathbb{V}, \mathbb{E}, \mathbb{Y}$ represent the set of nodes, edges, and labels, respectively.  
${\bf A} \in \{0,1\}^{|\mathbb{V}|\times |\mathbb{V}|}$ is the binary adjacency matrix of $G$, where ${\bf A}[v_j,v_k]=1$ if there exists an edge between nodes $v_j$ and $v_k$, and $0$ otherwise, with $|\mathbb{V}|$ the total number of nodes. 
FedGL employs FL techniques~\citep{mcmahan2017communication} to collaboratively train GL models with a set of (e.g., $T$) clients $\mathbb{T} = \{1, \cdots, T\}$ and a server. Assuming each client $i \in \mathbb{T}$ has a set of graphs $\mathbb{G}^i$, 
we illustrate the training process of FedGL using the $e$-th epoch as an example: 
1) Initially, the server distributes the global model parameters $\theta_e$ to a randomly selected subset of clients $\mathbb{T}_e$, where $\mathbb{T}_e \subseteq \mathbb{T}$.
2) Upon receiving $\theta_e$, each client $i$ trains its local model parameter $\theta^{i}_e$ with its own graphs $\mathbb{G}^i$ and updates its model parameters via SGD, i.e., $\theta^{i}_e = \theta^{i}_{e-1} - \eta \partial_{{\theta}_e} L({\theta}_e;\mathbb{G}^i)$, where $L({\theta}_e;\mathbb{G}^i)$ represents a loss function, e.g., cross-entropy loss. After training, client $i$ submits its update model parameters $\theta^{i}_e$ to the server.
3) The server aggregates local model parameters of the selected clients i.e., $\{\theta^{i}_e: i \in \mathbb{T}_e\}$ and updates the global model parameter, e.g., $\theta_{e+1} =\frac{1}{|\mathbb{T}_e|}  {\textstyle \sum_{i \in \mathbb{T}_e}} \theta^{i}_e$ via the average aggregation~\citep{mcmahan2017communication}, for the next epoch. 
This iterative process continues until the global model converges or reaches the maximum number of epochs.

\subsection{Backdoor-based Watermarking for GL}
\vspace{-2mm}
Backdoor-based watermarking~\citep{uchida2017embedding,adi2018turning, bansal2022certified,xu2023watermarking} adopts the idea of 
backdoor attack~\citep{bagdasaryan2020backdoor, wang2020attack, saha2020hidden} from the adversarial realm to  
facilitate model ownership verification. 
To watermark the GL model, assume the model owner has a set of clean graphs $\mathbb{G}$ and selects a subset of graphs $\mathbb{G}_w \subset \mathbb{G}$ to inject the watermark. 
In the existing method~\citep{xu2023watermarking}, the model owner 
first generates a random graph (e.g., via the ER-model) as the watermark for each to-be-watermarked graph. For instance, for a graph $G \in \mathbb{G}_w$ with label $y$, the generated random graph is $G_s$ (its size is often smaller than $G$). 
The owner then attaches $G_s$ to $G$ to produce the watermarked graph $G_w$, where nodes in $G$ are randomly chosen and the edge status of these nodes are replaced by edges in $G_s$.  
Finally, the owner assigns a desired label different from $y$ to 
$G_w$. 
The watermarked graphs together with the clean graphs are used to train the GL model. 
During model ownership verification, the one who can predict a high accuracy on these watermarked graphs can claim to be the model owner. 

We note this method can be extended to watermark FedGL models, where each client can generate its own random graphs as the watermark and train its local model with the watermarked graphs and clean graphs. The server then aggregates the watermarked local models to update the global model. 

\subsection{Watermark Removal Attacks}
\label{sec:threatmodel}
We consider three possible watermark removal attacks aiming to infringe the FedGL model ownership: 
distillation and finetuning from~\citep{shafieinejad2021robustness}, and our proposed layer-perturbation attack. 
In all attacks, the attacker (e.g., malicious user) is assumed to know the target watermarked model.  

\noindent {\bf 1) Distillation.} This attack has access to 
some unlabeled data sampled from the same data distribution. To remove watermarks without affecting the target model's main task performance, the attacker uses the unlabeled data to distill the target model during training. Specifically, 
{the attacker initializes its model with the target model and  
labels the unlabeled data by querying the target model. The attacker's model is then updated with these unlabeled data and their predicted labels. 
}

\noindent {\bf 2) Finetuning.} This attack assumes the attacker has some labeled data. The attacker then leverages the labeled data to further finetune the  
target model in order to forget the watermark.  
This attack is shown to pose a greater threat than the distillation attack~\citep{bansal2022certified}.  

\noindent {\bf 3) Layer-perturbation attack.} This attack also assumes the attacker has some labeled data. As knowing the target watermarked model (and hence the architecture), the attacker can mimic training an unwatermarked model with the same architecture as the target model using the labeled data. 
To further test the model robustness, we assume the attacker also knows some true watermarked samples, similar to~\citep{jiang2023ipcert}. 
Then, the attacker can replace \emph{any} layer(s)' parameters of the target model with those from the unwatermarked model to maximally reduce the watermark accuracy on its watermarked samples,  
while maintaining the main task performance. Our results (e.g., in Table~\ref{table:random-watermark}) show this layer-perturbation attack (even only perturbing 1 layer parameters) is much more effective than the other two attacks (even though the whole model parameters can be perturbed). 

\subsection{Threat Model}

We follow existing methods~\citep{shafieinejad2021robustness,bansal2022certified,xu2023watermarking,jiang2023ipcert}, where {the adversary is assumed to know all details of the pretrained watermarked FedGL model, but does not tamper with the training process.} This means all clients and the server are benign and follow the federated training protocol, and the attack happens at the testing/inference time.
We highlight this is in stark contrast to the training-time Byzantine attack on FL where some clients are malicious and they manipulate the training process.

\noindent {\bf {Attacker’s knowledge.}} The attacker has white-box access to the pretrained watermarked FedGL model. In addition, the attacker may also know some clean (unlabeled or labeled) training data, as well as watermarked data. Note that this setting actually makes our defense design the most challenging. If the defense can successfully defend against the strongest white-box attack on the watermarked FedGL model, it will also be effective against weaker attacks, such as black-box attacks.

\noindent {\bf {Attacker’s capability.}} The attacker can modify the pretrained model via leveraging its white-box access to the trained model and its hold training and watermarked data. For instance, the attacker can finetune the pretrained model via the labeled training data. More details of the capabilities of considered attacks are described in Section~\ref{sec:threatmodel}. 

\noindent {\bf {Attacker’s goal.}} The attacker aims to remove the watermark based on its knowledge and capability, while maintaining the model utility. This allows it to illegally use the model without detection.

\section{ {\name}: Our Certified Robust Watermark for FedGL}

\subsection{Motivation and Overview} 
\label{Motivation}
\vspace{-2mm}
Recall that  
the random graph based watermark is unable to ensure high watermark accuracy for protecting FedGL (as shown in Table~\ref{table:random-watermark}). This is because such random watermark does not use any graph structure  or client information during FedGL training. 
Our results also show this method is vulnerable to the three watermark removal attacks. 
These weaknesses inspire us to design a more effective and robust watermarking method specially for FedGL model ownership verification. 

We propose {\name}, the first certified robust backdoor-based watermarking method for FedGL. 
{\name} comprises two main components: \emph{Customized Watermark Generator (CWG)} and \emph{Robust Model Loader (RML)} (as depicted in Figure~\ref{fig:overview}). 
The CWG module utilizes the unique property of each client, as  different clients could have different properties (e.g., distributions of their graph data) and their optimal watermark could be different.
Particularly, CWG 
learns the customized watermark for each graph using its structure information, and outputs a set of diversified watermarked graphs for each client.  
Further, inspired by existing GNNs~\citep{xu2018powerful}, the RML module designs a new GL model that consists of multiple submodels,  each being any existing GL model. 
It also introduces a voting classifier for aggregating the prediction results from the submodels. Under this design, {\name} can be proved to be certified robust against the \emph{worst-case} layer-perturbation attack, once the number of perturbed layers is bounded.  
The model owner (e.g., participating clients in FedGL) adopts the designed GL model to train the local watermarked model with the learnt watermarked graphs and the remaining clean graphs. After the server-client training terminates, the ownership of the trained FedGL model can be verified via measuring its performance on a set of testing graphs injected with the \emph{global watermark}, which is 
the integration of all clients' local watermarks.  

\subsection{Customized Watermark Generator (CWG)} 
\vspace{-2mm}
CWG consists of two networks: $\mathrm{GatingNet}$ and $\mathrm{KeyNet}$. $\mathrm{GatingNet}$ designs the watermark for each graph separately using the edge information, 
while $\mathrm{KeyNet}$ learns client-wise  watermarking style using predefined keys (e.g., client ID in this paper). 
The customized watermark for each client's graph is then decided using the output of  $\mathrm{GatingNet}$ and $\mathrm{KeyNet}$.  
\emph{Detailed network architectures of CWG can be seen in 
Table~\ref{app:CWG} in Appendix~\ref{app:experiment setup}.} We demonstrate how CWG can learn a customized  
watermark using a graph $G^i= (\mathbb{V}^i,\mathbb{E}^i)$ from client $i$ as an instance. The details  
are as follows:
\begin{itemize}[leftmargin=*]
\vspace{-2mm}
\item We first randomly select $n_w$ nodes $\mathbb{V}_w^i=\{v_1, \cdots, v_{n_w}\}$ from $\mathbb{V}^i$ as watermark nodes and construct a corresponding mask matrix ${\bf M}^i \in \{0,1\}^{|\mathbb{V}^i|\times |\mathbb{V}^i|}$ such that ${\bf M}^i[v_j,v_k]=1$ if $v_j,v_k \in \mathbb{V}_w^i$, and  $0$ otherwise. We also update the adjacency matrix ${\bf A}^i$ of $G^i$
according to $\mathbb{V}_w^i$, i.e., setting ${\bf A}^i[v_j,v_k]=0, \forall v_j,v_k \in \mathbb{V}_w^i$. This allows us focus on learning the edge status between watermarked nodes. 
\vspace{-1mm}
\item  
Given the client $i$'s  ID string $k^i$,  we utilize a cryptographic hash function, such as MD5, to convert it into an integer (e.g., $128$-bit long with the integer range $[0, 2^{128} - 1]$). This integer is then employed as a seed to produce a  key matrix ${\bf K}^i \in \mathbb{R}^{|\mathbb{V}^i|\times |\mathbb{V}^i|}$. Then, we employ $\mathrm{GatingNet}$ and $\mathrm{KeyNet}$ to extract edge features and  key features, resulting in ${\tilde{\bf A}}^i =\mathrm{GatingNet}({\bf A}^i) \in [0,1]^{|\mathbb{V}^i|\times |\mathbb{V}^i|}$ and ${\tilde{\bf K}}^i = \mathrm{KeyNet}({\bf K}^i) \in [0,1]^{|\mathbb{V}^i|\times |\mathbb{V}^i|}$, respectively.  

\vspace{-1mm}
\item We finally learn the customized watermark for $G^i$ by 
integrating 
${\tilde{\bf A}}^i$, ${\tilde{\bf K}}^i$, and ${\bf M}^i$, and obtain the corresponding watermarked graph as $G_w^i = (\mathbb{V}^i,\mathbb{E}_w^i)$. Here $\mathbb{E}_w^i$ is the set of edges according to the updated adjacency matrix ${\bf A}^i\oplus {\bf W}^i$, where $\oplus$ is the element-wise addition and ${\bf W}^i  = \mathbb{I}(({\tilde{\bf A}}^i \odot {\tilde{\bf K}}^i)>0.5) \odot  {\bf M}^i$ contains the edge status between the watermark nodes $\mathbb{V}_w^i$. Here, $\odot$ is the element-wise product, $\mathbb{I}(p)$ is an indicator function returning $1$ if $p$ is true, and $0$ otherwise. 
We adopt $0.5$ as a threshold to decide the presence of edges between watermarked nodes. 
\end{itemize}

\subsection{Robust Model Loader (RML)} 
\vspace{-2mm}
This module aims to design a new GL model that is provably robust to the layer-perturbation attack. Towards this end, 
we design  
a GL model architecture to incorporate multiple submodels; and devise a majority voting-based ensemble classifier 
on top of the predictions of these submodels. 

\noindent {\textbf{Architecture of the proposed GL model.}} 
{Intuitively, each client can take a base GL model (e.g., GIN~\citep{xu2018powerful})  
and split it according to the layer indexes to obtain multiple submodels.  
For instance, a $8$-layer GIN can be represented with layer indexes $\{ l_1, \cdots, l_8\}$. Splitting this GIN into $4$ submodels $\{$GIN$_1, \cdots,$ GIN$_4\}$ with layer indexes $\{l_1,l_2\}, \cdots, \{l_7,l_8\}$ means GIN$_i$ contains layers $\{l_{2i-1},l_{2i}\}$, from the GIN. {However, submodels splitted in this way are coupled from each other, making them unable to defend against layer-perturbation attacks.} To tackle this problem, we design the novel GL model $\theta$ that is an ensemble of a set of $S$ \emph{independent} submodels $\{\vartheta_1, \vartheta_2, \cdots, \vartheta_S\}$, where each submodel $\vartheta_i$ is a base GL model. This approach can hence be easily adapted to any existing FedGL.} Further, to prevent homogeneity,  
we define varying channels for these submodels to diversify them. Details of the model architecture are shown in Table~\ref{app:model} in Appendix~\ref{app:experiment setup}.

\noindent {\bf A majority-voting based ensemble classifier.}  
The designed GL model architecture inspires 
us to leverage the idea of ensemble classifier, which can combine the predictions of base ``weak'' classifiers. Specifically, we propose a majority voting-based ensemble classifier to combine the predictions of the submodels. 
Given a testing graph $G$ and a graph classifier $f$, we denote the prediction of the submodel $\vartheta_i$ for  $G$ as $y = f(\vartheta_i, G) \in \mathbb{Y}$. 
For a GL model $\theta$ with $S$ submodels $\{\vartheta_1, \vartheta_2, \cdots, \vartheta_S\}$, we can 
count the submodels that classify $G$ to be $y$ as $N_y={\textstyle \sum_{i=1}^S} \mathbbm{I}(f(\vartheta_{i},G) = y)$.  
Then we introduce our majority-voting based ensemble classifier $g$ to classify $G$ as: $g(\theta,G)={\arg\max}_{y \in \mathbb{Y}} \ N_y$. 
In cases of ties, our ensemble classifier $g$ selects the label with a smaller index.

\vspace{-2mm}
\subsection{Training the Proposed FedGL Model}
\vspace{-2mm} 
The overall training process consists of three iterative steps: 1) training the proposed GL model in all clients; 2) training the CWG module in \emph{watermarked clients}, i.e., the clients that aim to inject watermarked graphs for protecting the model ownership; 
and 3) aggregating the clients' GL models to produce the target watermarked model.  
The final global model is the learnt watermarked FedGL model. 
Details of training can be seen in  Algorithm~\ref{algorithm: training} in the Appendix. 

{\bf Step 1: Training the proposed GL model.} Assume we have $T_w$ watermarked clients with indexes $[1,T_w]$. For each watermarked client $i$, we split its training graphs $\mathbb{G}^i$ into the watermarked graphs $\mathbb{G}^i_w$ with a target label, say $y_w$, and remaining clean graphs $\mathbb{G}^i_c$, and then customize the watermark for each graph in $\mathbb{G}_w^i$ using the $\mathrm{CWG}$ module (see {\bf Step 2}). Given the client's GL model $\theta^i$ with $S$ submodels $\{\vartheta^i_1, \vartheta^i_2, \cdots, \vartheta^i_S\}$, we train each submodel $\vartheta^i_j$ via minimizing the loss on $\mathbb{G}^i_c$ and $\mathbb{G}^i_w$,  
i.e., $\vartheta^i_j = {\arg\min}_{\vartheta^i_j} \, L(\vartheta^i_j;\mathbb{G}^i_c \bigcup \mathbb{G}^i_w)$. 
For an unwatermarked client $k \in [T_w+1, T]$, we utilize all clean graphs $\mathbb{G}^k$ to train each submodel $\vartheta^k_j$ separately, i.e., $\vartheta^k_j = {\arg\min}_{\vartheta^k_j} \, L(\vartheta^k_j;\mathbb{G}^k)$. 

{\bf Step 2: Training the CWG.} We denote the parameters of the CWG module for a watermarked client $i \in [1,T_w]$,  
as $\omega^i$. The parameters include two networks, $\mathrm{GatingNet}$ and $\mathrm{KeyNet}$. 
Each client $i$ trains its CWG $\omega^i$ to ensure that the generated watermarks be effective and diverse. Formally, we have $\omega^{i} = {\arg\min}_{\omega^{i}} \, L(\theta^{i}; \mathbb{G}^i_{w})$, $i \in [1, T_w]$. 

{\bf Step 3: Aggregating clients' GL models.} The server averages GL models $\{\theta^i\}_{i \in T}$ 
to produce the global model $\theta$, and distributes this model to selected clients in the next iteration.

\subsection{Model Ownership Verification}
\vspace{-2mm}
When suspecting the target FedGMark model $\theta$ is illegally used by others, the model owner (all the participating clients or their representative) can recruit a trusted judge for model ownership verification.   
Typically, the judge requests both the true model owner and the illegal party to provide some test data for verification.  
Only when the one knows the predictions by the target model for the provided test data by both parties, the judge will confirm this party the model ownership.  In particular, besides providing the clean data $\mathbb{G}^i_c$ by both parties that behave normally, the true model owner especially provides the designed watermarked data $\mathbb{G}^i_w$ that only s/he knows the model behaves on. As a result, both parties know the prediction results on $\mathbb{G}^i_c$, but the illegal party is hard to predict accurately on $\mathbb{G}^i_w$ provided by the true model owner.

\vspace{-2mm}
\subsection{Certified Robustness Guarantees against Layer-Perturbation Attacks}
\vspace{-2mm}
We show the above design, with any layer-perturbation attack,  ensures the predictions of the learnt  watermarked FedGL model  
and its compromised counterpart for the watermarked graphs are  consistent, once the number of perturbed layers is bounded. 
Given the target watermarked FedGL model $\theta$ and its $S$ submodels $\{\vartheta_1, \cdots, \vartheta_S\}$,  we denote $\theta'$ as the comprised model  
and 
$\{\vartheta_1', \cdots, \vartheta_S'\}$ as its $S$ submodels. 
For each watermarked graph $G_w$,  we use the ensemble classifier $g$ on submodels' predictions, i.e.,  
its predictions on $\theta$ and $\theta'$ are 
$g(\theta,G_w)={\arg\max}_{y \in \mathbb{Y}} N_y$, and $g(\theta',G_w)={\arg\max}_{y \in \mathbb{Y}} N_y'$, 
respectively, 
where $N_y={\textstyle \sum_{i=1}^S} \mathbbm{I}(f(\vartheta_{i},G) = y)$  
and $N'_y={\textstyle \sum_{i=1}^S} \mathbbm{I}(f(\vartheta_{i}',G) = y)$.
Then we have the following result on guaranteeing the number of perturbed layers on the target watermarked model. 

\begin{theorem}[Certified number of perturbed layers $r$.]
\label{thm:certified} 
\vspace{-2mm}
Let  $\theta$, $\theta'$, $g$, and $G_w$ be above defined. 
Suppose $N_A$ and $N_B$ are the largest and second largest count outputted by  
$g$ on $G_w$, For any layer-perturbation attack, 
we have $g(\theta,G_w) = g(\theta',G_w)$, when the number of perturbed layers $r$ satisfies: 
\begin{equation}
r \le  r^* = \left(N_A - N_B + \mathbbm{I}[A<B] -1 \right) \big / 2,
\label{eq: thm1}
\end{equation}
where $\mathbbm{I}[\cdot]$ is the indicator function and $r$ is called the certified number of perturbed layers. 
\label{thm:certify}
\vspace{-2mm}
\end{theorem}

We also show the tightness of our derived $r^*$ in the following theorem:
\vspace{-2mm}
\begin{theorem}[Tightness of $r^*$.]
\label{thm:tight}
Without using extra information of $f$, our derived $r^*$ in Theorem~\ref{thm:certified}  is tight. I.e., 
$r^*$ is the maximum number of perturbed layers tolerated by our target watermarked model. 
\end{theorem}
\vspace{-2mm}
The {proofs} of Theorems \ref{thm:certify} and \ref{thm:tight} are deferred to Appendix~\ref{app:thmproofs}.

\section{Experiments}
\vspace{-2mm}
In this section, we comprehensively evaluate {\name} 
on multiple datasets, FedGL models, attack baselines, and experimental settings. More experimental results and discussions are in Appendix. 

\vspace{-2mm}
\subsection{Experimental Setup}
\label{Setup}
\vspace{-2mm}
\noindent {\bf Datasets and models.} We evaluate our {\name} on four real-world graph datasets for graph classification: MUTAG~\citep{debnath1991structure}, PROTEINS~\citep{borgwardt2005protein}, DD~\citep{dobson2003distinguishing}, and COLLAB~\citep{yanardag2015deep}. Details about the statistics of those datasets are shown in Table~\ref{table:datasets} in Appendix~\ref{app:experiment setup}. Following prior work~\cite{shen2022model,yang2023graphguard}, we choose the well-known GIN~\citep{xu2018powerful}, GSAGE~\citep{hamilton2017inductive}, and GCN~\citep{kipf2016semi} as the GL model.  
All these network architectures involved in the experiments are detailed in Table~\ref{app:model} in Appendix~\ref{app:experiment setup}. 

\vspace{-1mm}
\noindent {\bf Parameter setting.} We implement our method using one NVIDIA GeForce GTX 1080 Ti GPU. In FedGL, we use $T=40$ clients in total and train the model 200 iterations. The server randomly selects $50\%$ clients 
in each iteration. 
We define the target label of watermarking graphs as $1$, and each participating client randomly selects $10\%$ graphs with labels not $1$ as the watermarking graphs. We extensively validate the effectiveness and robustness of the proposed watermarking method with the following hyperparameters details: the number of submodels $S =\{4,8,16\}$, the number watermarked clients $T_w=\{5,10,20\}$ (both $S$ and $T_w$ are halved on MUTAG due to less data), the watermarked nodes $n_w=\{3,4,5\}$, and the number of perturbed layers $r=\{1,\cdots, 5\}$ in the layer-perturbation attack. By default, we set $S=4$, $T_w=10$, $n_w=4$, $r=1$. 
While studying the impact of a hyperparameter, we fix the others as the default value. 

\vspace{-1mm}
\noindent {\bf Evaluation metric.} We use three metrics for evaluation:  
the main task accuracy (MA), watermark accuracy (WA), and certified WA (CWA$@r$). 
An effective and robust watermarked model is expected to achieve both high MA and WA. 
CWA  evaluates the certified robustness of {\name} against layer-perturbation attacks. CWA$@r$ is defined as the fraction of testing graphs that are provably predicted as the target label, when at most $r$ layers in client models can be arbitrarily perturbed. 

\vspace{-1mm}
\noindent {\bf {Attack baselines.}} 
We evaluate FedGMark against existing  
watermark removal attacks including distillation and finetuning, and our proposed layer-perturbation attack. In our setting, the layer-perturbation attack can replace \emph{any} layer(s)' parameters of the target watermarked model with those from the unwatermarked model to maximally reduce the watermark accuracy on its watermarked graphs. When perturbing multiple layers,  
we utilize a greedy algorithm to decide the  perturbed layers---we search 
for one optimal perturbed layer at each step. Specifically, we first traverse perturbing layers in the watermarked model and find the one that maximally reduces the watermark accuracy. We then search for the remaining layers based on this one and continue the process.  

\begin{table}[!t]
\scriptsize
\renewcommand{\arraystretch}{1.1}
\addtolength{\tabcolsep}{-4pt}
\caption{Results of our {\name} under empirical watermark removal attacks. 
}
\centering
    \begin{tabular}{c||c|c c c c||c|c c c}
    \toprule
    {Datasets}  & \multicolumn{1}{c}{Attack}&  {Fed-GIN}  &  {Fed-GSAGE} & {Fed-GCN} &{Datasets}  & \multicolumn{1}{c}{Attack}&  {Fed-GIN}  &  {Fed-GSAGE} & {Fed-GCN}  \\ 
    & & MA$\uparrow$ \ WA$\uparrow$ & MA$\uparrow$ \ WA$\uparrow$  & MA$\uparrow$ \ WA$\uparrow$  & & & MA$\uparrow$ \ WA$\uparrow$  & MA$\uparrow$ \ WA$\uparrow$  & MA$\uparrow$ \ WA$\uparrow$ \\
    \hline
    \hline
    & None & {\bf 0.81 \ \ 0.90} & {\bf 0.85 \ \ 0.89} & {\bf 0.83 \ \ 0.83} & &None & {\bf 0.72 \ \ 0.86} & {\bf 0.72 \ \ 0.82} & {\bf 0.73 \ \ 0.84} \\
    MUTAG &  Distillation & 0.83 \ \ 0.87  & 0.81 \ \ 0.70  & 0.80 \ \ 0.80 &  PROTEINS&Distillation & 0.70 \ \ 0.84 & 0.70 \ \ 0.80 & 0.70 \ \ 0.79\\
    & Finetuning&  0.82 \ \ 0.88  & 0.79 \ \ 0.76  & 0.78 \ \ 0.81 &  & Finetuning &  0.70 \ \ 0.84   & 0.71 \ \ 0.82 & 0.72 \ \ 0.80\\ 
    & 1-Layer Pert.  & {\bf 0.82 \ \ 0.90} & {\bf 0.83 \ \ 0.91} & {\bf 0.81 \ \ 0.84} & & 1-Layer Pert.  & {\bf 0.73 \ \ 0.86} & {\bf 0.74 \ \ 0.83} & {\bf 0.74 \ \ 0.85}\\
    \hline 
    & None &{\bf 0.73 \ \ 0.65}& {\bf 0.74 \ \ 0.57}  & {\bf 0.74 \ \ 0.56} & & None & {\bf 0.73 \ \ 0.75} & {\bf 0.74 \ \ 0.74} & {\bf 0.73 \ \ 0.70}\\
    DD &  Distillation & 0.71 \ \ 0.63  & 0.74 \ \ 0.56 & 0.72 \ \ 0.56 & COLLAB & Distillation & 0.73 \ \ 0.71 & 0.72 \ \ 0.73 & 0.70 \ \ 0.69 \\
    & Finetuning& 0.72 \ \ 0.62  & 0.74 \ \ 0.57 & 0.74 \ \ 0.55 &  & Finetuning & 0.72 \ \ 0.70 & 0.75 \ \ 0.71 & 0.72 \ \ 0.67 \\   
    & 1-Layer Pert.  &  {\bf 0.72 \ \ 0.65}  & {\bf 0.75 \ \ 0.55} & {\bf 0.73 \ \ 0.57} & & 1-Layer Pert.  & {\bf 0.72 \ \ 0.76} & {\bf 0.73 \ \ 0.74} & {\bf 0.74 \ \ 0.71} \\
    \bottomrule
    \end{tabular}
\label{table:experiment-total}  
\end{table}

\begin{figure*}[!t]
\centering	
\includegraphics[width=\textwidth]{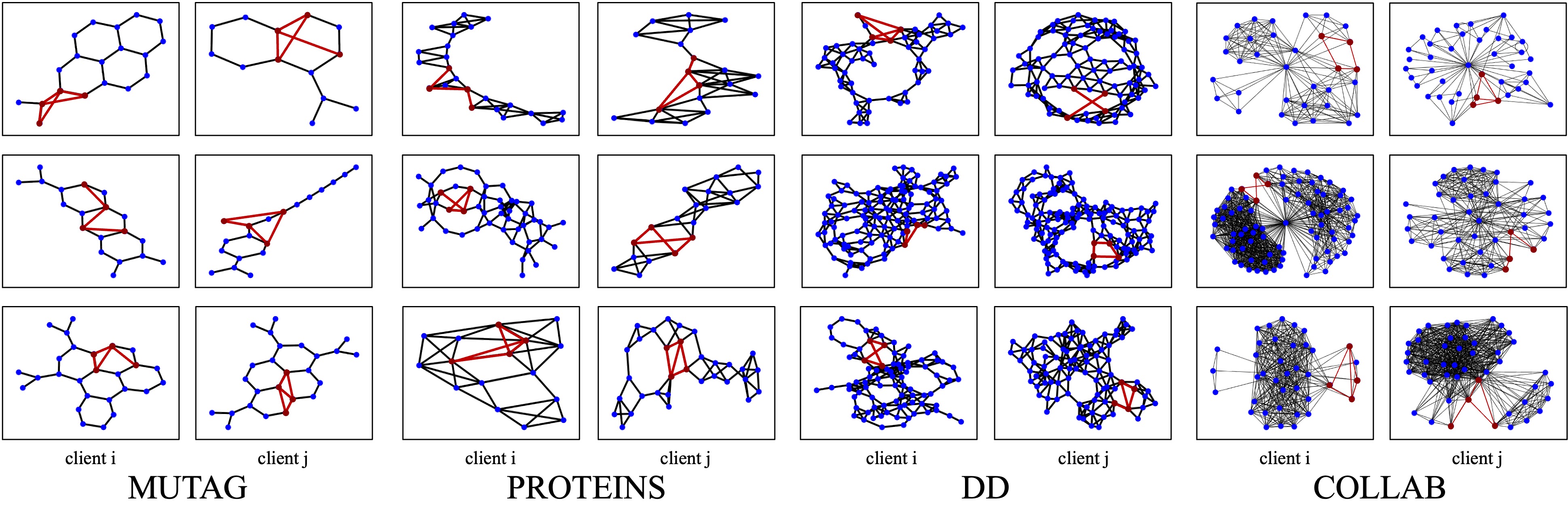}
    \vspace{-7mm}
	\caption{Example learnt watermarks and watermarked graphs by our {\name}. CWGs generated by different clients produce unique watermarks, characterized by distinct edge connection patterns.} 
        \label{fig:watermarked graphs}
         \vspace{-1mm}
\end{figure*}

\begin{table}[!t]
\scriptsize
\renewcommand{\arraystretch}{1.1}
\addtolength{\tabcolsep}{-4.2pt}
\caption{Impact of $S$ on {\name} against our layer-perturbation attack.
}
\centering
    \begin{tabular}{c||c|c c c c||c|c c c}
    \toprule
    {Datasets}  & \multicolumn{1}{c}{Attack}&  {Fed-GIN}  &  {Fed-GSAGE} & {Fed-GCN} &{Datasets}  & \multicolumn{1}{c}{Attack}&  {Fed-GIN}  &  {Fed-GSAGE} & {Fed-GCN}  \\ 
    ($S = 8$)& & MA$\uparrow$ \ WA$\uparrow$ & MA$\uparrow$ \ WA$\uparrow$  & MA$\uparrow$ \ WA$\uparrow$  & & & MA$\uparrow$ \ WA$\uparrow$  & MA$\uparrow$ \ WA$\uparrow$  & MA$\uparrow$ \ WA$\uparrow$ \\
    \hline
    \hline
    MUTAG & None & {\bf  0.83\ \ 0.93 } & {\bf 0.83\ \ 0.92 } & {\bf 0.78\ \ 0.87} & PROTEINS &None & {\bf 0.73\ \ 0.86 } & {\bf 0.71\ \ 0.85} & {\bf 0.74\ \ 0.88} \\ 
    & 1-Layer Pert.  & {  0.79\ \ 0.93} & {  0.84\ \ 0.92} & {  0.79\ \ 0.76 } & & 1-Layer Pert.  & {  0.70\ \ 0.86} & {  0.71\ \ 0.85} & {  0.73\ \ 0.88}\\
    \hline
     DD& None & {\bf 0.71\ \ 0.57 } & {\bf 0.73\ \ 0.52} & {\bf 0.74\ \ 0.50} & COLLAB &None & {\bf 0.73 \ \ 0.73 } & {\bf 0.74 \ \ 0.74 } & {\bf  0.72 \ \ 0.72} \\ 
    & 1-Layer Pert.  & {  0.72\ \ 0.56} & {  0.73\ \ 0.50} & {  0.74\ \ 0.50} & & 1-Layer Pert.  & {   0.73 \ \ 0.72} & {  0.74 \ \ 0.75 } & {  0.72 \ \ 0.72 }\\
    \hline
    \hline
    ($S= 16$)& None & {\bf 0.83\ \ 0.90} & {\bf  0.86\ \ 0.87} & {\bf 0.79\ \ 0.79} & PROTEINS &None & {\bf 0.72\ \ 0.86} & {\bf 0.72\ \ 0.85} & {\bf 0.71\ \ 0.85} \\ 
    MUTAG & 1-Layer Pert.  & {  0.81\ \ 0.90} & {  0.84\ \ 0.86} & {  0.79\ \ 0.80 } & & 1-Layer Pert.  & {  0.73\ \ 0.86} & {  0.72\ \ 0.85} & {  0.73\ \ 0.85}\\
    \hline
    DD & None & {\bf 0.70\ \ 0.61} & {\bf 0.74\ \ 0.54} & {\bf 0.73\ \ 0.50} &COLLAB &None & {\bf 0.71 \ \ 0.76 } & {\bf 0.72 \ \ 0.74 } & {\bf 0.72 \ \ 0.72 } \\ 
    & 1-Layer Pert.  & {  0.73\ \ 0.61} & {  0.74\ \ 0.54} & {  0.74\ \ 0.50 } & & 1-Layer Pert.  & {  0.71 \ \ 0.76 } & {  0.71 \ \ 0.73 } & {  0.72 \ \ 0.72 }\\
    \bottomrule
    \end{tabular}
\vspace{-3mm}
\label{table:experiment-total-8-and-16}  
\end{table}

\subsection{Empirical Results: MA and WA}

\subsubsection{Results under the Default Setting}
\vspace{-2mm}
We first assess our {\name} against empirical and layer-perturbation attacks 
under the default setting. 
Experimental results are provided in Table~\ref{table:experiment-total}. 
We have the following observations: 
\begin{itemize}[leftmargin=*]
\vspace{-2mm}
\item {\bf 1) Our {\name} significantly outperforms the existing method under no attack.} Recall in Table \ref{table:random-watermark} that 
\citep{xu2023watermarking} obtains $<60\%$ WAs across all datasets and FedGL models. 
In contrast, {\name} can achieve WAs $>70\%$ in almost all cases, while having similar MAs. 
Such significant 
improvements confirm the superiority of our learnt watermarks over the random watermarks in \citep{xu2023watermarking}. Figure~\ref{fig:watermarked graphs}  
also visualizes example learnt watermarks and we can see these watermarks are diversified due to the proposed CWG.

\item {\bf 2) Our {\name} exhibits resistance 
to existing empirical attacks.} As shown in Table \ref{table:random-watermark}, existing methods are vulnerable to distillation and finetuning attacks. In contrast, the WAs of {\name} under these two attacks are almost the same as those under no attack, demonstrating {\name} is robust to the existing attacks. 

\item {\bf 3) Our {\name} is resilient to 
the proposed layer perturbation attack.} 
We notice that the existing method  has difficulties in defending against the proposed layer-perturbation attacks and shows an unsatisfactory WA, e.g., $<25\%$ WA in almost all cases under the 1-layer perturbation attack. 
Conversely, our {\name}  
can obtain a close WA  
compared to that without attack.  
This verifies our ensemble classifier in {\name} is capable of resisting the 1-layer perturbation attack.

\end{itemize}

\vspace{-2mm}
\subsubsection{Impact of Hyperparameters}
\vspace{-2mm}
This section studies the impact of hyperparameters in {\name} against the layer-perturbation attack.  

{\bf Impact of \#submodels $S$.} We first examine the impact of 
$S$ on the performance of {\name} and report the results against the 1-layer perturbation attack in Table~\ref{table:experiment-total-8-and-16}. We observe that {the learnt watermarked model can resist to the 1-layer perturbation attack under all $S$} and both the MAs and WAs are also similar. 
This indicates the number of submodels marginally affects the watermarking performance against the 1-layer perturbation attack.

\begin{table}[!t]
\scriptsize
\renewcommand{\arraystretch}{1.1}
\addtolength{\tabcolsep}{-3.2pt}
\caption{Impact of \#perturbed layers on FedGMark against our layer-perturbation attack.   
Compared with Table~\ref{table:experiment-total}, the change in MA is less than $4\%$ in all cases.
} 
\centering
    \begin{tabular}{c||c|c c c c c|c||c|c c c c c}
    \toprule
    {Datasets}  & \#Perturbed Layers&  1  &  2 & 3 & 4 & 5 &{Datasets}  & \#Perturbed Layers&  1  &  2 & 3 & 4 & 5 \\ 
    & (Net) & WA$\uparrow$ & WA$\uparrow$  & WA$\uparrow$ & WA$\uparrow$  & WA$\uparrow$ & & (Net) & WA$\uparrow$ & WA$\uparrow$  & WA$\uparrow$ & WA$\uparrow$  & WA$\uparrow$ \\
    \hline
    & Fed-GIN & 0.90  & 0.89  & 0.54 & 0.23  &0.16  & &Fed-GIN &  0.86 &  0.84 & 0.58 &  0.46 & 0.39 \\
    MUTAG &  Fed-GSAGE & 0.91  &  0.89  & 0.55 &  0.21  & 0.13 & PROTEINS &Fed-GSAGE &  0.83 & 0.81 & 0.55 &  0.45 & 0.44 \\
    & Fed-GCN &  0.84 &  0.83  & 0.49 &  0.24  & 0.18 &  & Fed-GCN & 0.85  & 0.82 & 0.57 & 0.48 & 0.41 \\
    \hline
    & Fed-GIN &  0.65 & 0.61  & 0.56 &  0.32 & 0.21  & &Fed-GIN &  0.76 & 0.74 & 0.58 &  0.36 & 0.27 \\
    DD &  Fed-GSAGE & 0.55  &  0.55  & 0.49 &  0.28  & 0.24 & COLLAB &Fed-GSAGE & 0.74  &  0.73 & 0.57 & 0.35  & 0.25 \\
    & Fed-GCN &  0.57   &  0.54  & 0.48 & 0.29 & 0.24 &  & Fed-GCN & 0.71  & 0.70 & 0.52 & 0.32 & 0.21 \\
    \bottomrule
    \end{tabular}
\label{table:Attack levels vs. WA}  
\vspace{-4mm}
\end{table}

{\bf Impact of \#watermarking clients $T_w$.} 
The  results on  $T_w=5$ and $T_w=20$ are shown in Figure~\ref{fig:experiment-total-nc=5-and-20} in Appendix. We see that: \emph{the larger number of clients generating watermarks, the better our watermarking performance.} For instance, the WA  improves by $11\%$, $9\%$, and $7\%$, respectively on the three FedGL models on PROTEINS when $T_w$ increases from $5$ to $20$. This is because more watermark clients (thus more watermarked graphs) can ensure the FedGL model better learns the relation between the watermarks and the target label. 

{\bf Impact of the watermark size $n_w$.} We also  investigate $n_w = 3$ and $n_w = 5$, and the  results are presented in Figure~\ref{fig:experiment-total-ntri=3-and-5} in Appendix. Similarly, {\name} obtains higher WA with larger $n_w$. This is because a larger watermark size can facilitate the trained FedGL model better learn the watermark.

\begin{figure*}[!t]
\centering	
\includegraphics[width=\textwidth]{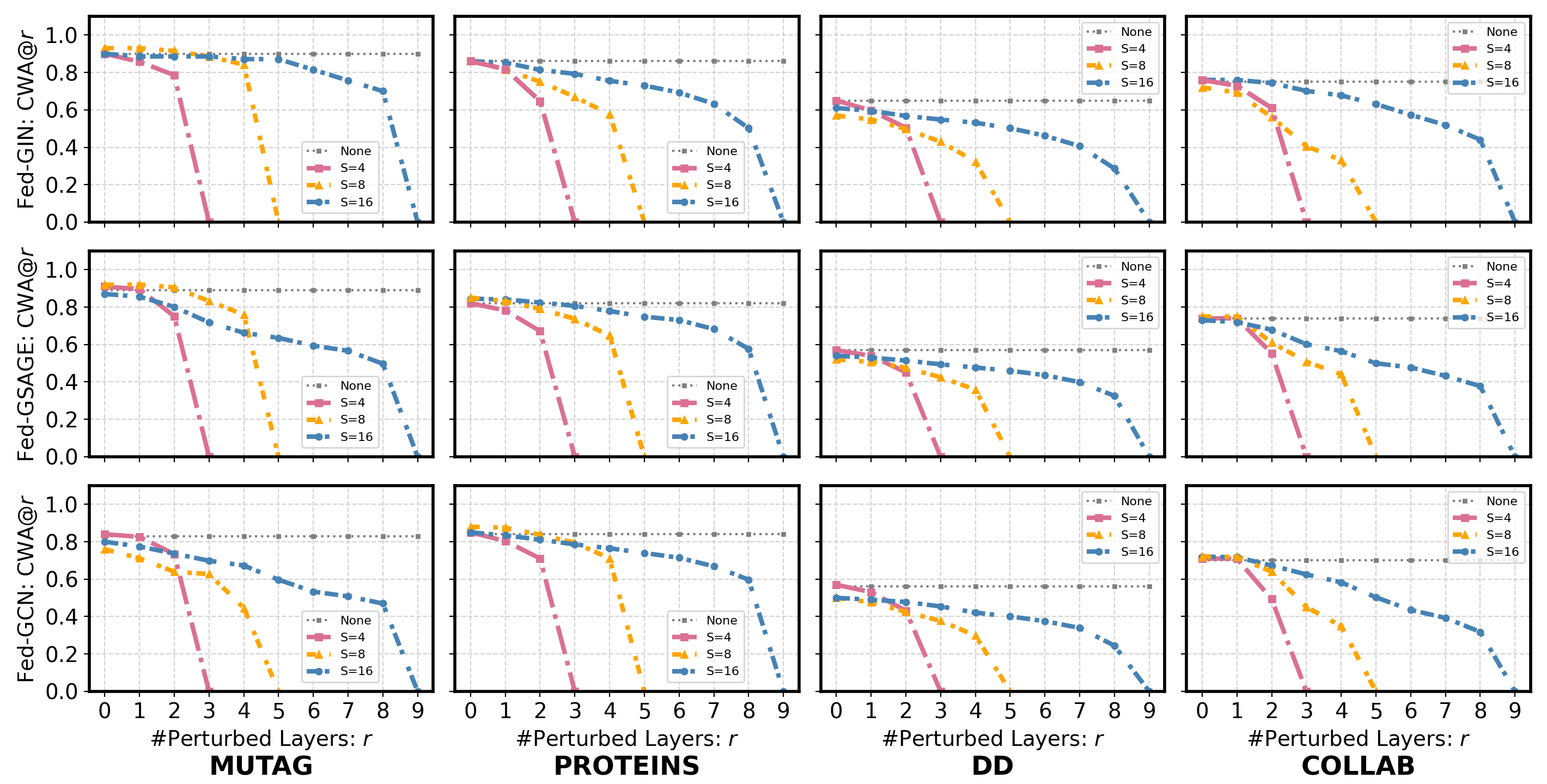}
    \vspace{-6mm}
	\caption{CWA vs. \#perturbed layers $r$ in the layer-perturbation attack. 
    }  
        \label{fig:certified-r}
    \vspace{-6mm}
\end{figure*}

{\bf Impact of \#perturbed layers.}  
Finally, Table~\ref{table:Attack levels vs. WA} shows the results of {\name} vs. different \#perturbed layers. We observe that:  
1) The WA decreases as increasing the \#perturbed layers. 
This is because the attacker has more attack capability by perturbing more layers. 
2) Despite the theoretically guaranteed CWA being $0$ at \#perturbed layers $\geq 3$ as later depicted in the Figure~\ref{fig:certified-r}, our empirical watermarking performance can still achieves WA from $48\%$ to $58\%$. 
Note that the change in MA is $<4\%$ in all cases, compared with Table~\ref{table:experiment-total}.

\begin{wraptable}{r}{0.57\textwidth}
\scriptsize
\renewcommand{\arraystretch}{1.0}
\addtolength{\tabcolsep}{-5pt}
\vspace{-8mm}
\caption{Impact of different modules in {\name}.
}
\centering
    \begin{tabular}{c||c c c|c c c c}
    \toprule
    {\bf Models} &  G.Net  &  K.Net & RML & MUTAG & PROTEINS & DD & COLLAB\\ 
    & & & & MA$\uparrow$ \ WA$\uparrow$ & MA$\uparrow$ \ WA$\uparrow$ & MA$\uparrow$ \ WA$\uparrow$ & MA$\uparrow$ \ WA$\uparrow$ \\
    \hline
    {\bf (a)} & \ding{51} & \ding{51} & \ding{51} & {\bf 0.81 \ \ 0.90} & {\bf 0.72 \ \ 0.86} & {\bf 0.73 \ \ 0.65} & {\bf 0.73 \ \ 0.75} \\
    {\bf (b)} & \ding{51} &  & \ding{51} & 0.81 \ \ 0.84 & 0.72 \ \ 0.80 & 0.72 \ \ 0.57 & 0.72 \ \ 0.70\\
    {\bf (c)} & & \ding{51} & \ding{51} & 0.80 \ \ 0.69 & 0.72 \ \ 0.68 &0.72 \ \ 0.48 & 0.73 \ \ 0.67 \\
    {\bf (d)} & \ding{51} & \ding{51} & & 0.81 \ \ 0.89 & 0.73 \ \ 0.86 & 0.71 \ \ 0.66 & 0.72 \ \ 0.75 \\
    {\bf (e)} & & & & 0.82 \ \ 0.47 & 0.71 \ \ 0.56 & 0.73 \ \ 0.39 & 0.73 \ \ 0.57 \\
    \bottomrule
    \end{tabular}
\label{table:ablation study} 
\vspace{-4mm}
\end{wraptable}

\subsubsection{Ablation Study} 
\vspace{-2mm}
We investigate the contribution of each module in FedGMark, including the $\mathrm{GatingNet}$ and $\mathrm{KeyNet}$ in CWG, and the submodels used in RML.  
The results under the default setting and no attack are shown in Table~\ref{table:ablation study}. 
First, $\mathrm{GatingNet}$ plays a crucial role on generating more effective watermarks and results in an improvement in WA from $8\%$ to  $21\%$. Second, $\mathrm{KeyNet}$ facilitates watermark differentiation between clients, leading to an improve on WA from 5\% to 8\%. 
Third,  
the submodels in RML has a negligible impact on  WA/MA. 
However, it is important to defend against the layer-perturbation attack, as shown in Section~\ref{sec:results:CWA}.

\vspace{-2mm}
\subsection{Certified Robustness Results: Certified WA}
\vspace{-2mm}
\label{sec:results:CWA}

In this section, we evaluate the certified robustness 
of {\name} against the  
layer perturbation attack. 
The  CWAs on the three FedGL models and four datasets 
are depicted in Figure~\ref{fig:certified-r}. 
We have several observations: 
{1) {\name} achieves promising provable robustness results against the \emph{worst-case} layer perturbation attack,  
when the \#perturbed layers is within the certified range in Eqn (\ref{eq: thm1}).} For instance, when $S=4$ and $0< r \leq 2$, CWA is close to 
WA without attack ($r=0$). 
{2) As 
$S$ increases, the certified \#perturbed layers also increases, showing that a more number of submodules in RML can better provably defend against the layer-perturbation attack.} 
For instance, when $S=16$, the CWA is $86\%$ on MUTAG even when 5 any layers in the global model 
are arbitrarily perturbed, while it is 0 when $S=4$. 
However, this is at the cost of requiring more computational resources (e.g., GPU memory, runtime). As shown in Table~\ref{table:The Cost of RML} in Appendix, the runtime is linearly to $S$.

\vspace{-2mm}
\section{Related Work}
\vspace{-2mm} 

{\bf Backdoor-based watermarking for centralized models on non-graph data.} Many backdoor-based watermarking methods~\citep{lv2023robustness,yan2023rethinking,tekgul2021waffle, yang2023watermarking, li2022fedipr, shao2022fedtracker, lansari2023federated} have been proposed that can \emph{empirically} protect the model ownership. These methods mainly focus on centralized learning models on non-graph (e.g., image) data.  
Compared with non-graph data, graph data have unique graph structure information, e.g., entities are connected by links. Similarly, compared with centralized models, FL models are collaboratively trained by multiple clients, which could have their own uniqueness. 
\cite{bansal2022certified} is the first watermarking method  for centralized non-graph models with certified guarantee. 
However, its certified robustness performance is unsatisfactory.

{\bf Backdoor-based watermarking for FL models on non-graph data.} A few recent works \cite{tekgul2021waffle,li2022fedipr,bansal2022certified}  design backdoor-based watermarks for protecting the ownership of FL models on non-graph data, where the watermark can be injected into client's data or server's validation data. 
For instance, \cite{tekgul2021waffle} presented WAFFLE, an approach to watermark DNN models by incorporating a re-training step via watermarked data at the server. 
\cite{li2022fedipr} leveraged each client's private watermark to verify FL model ownership, ensuring non-conflicting watermarks across different clients. 
However, these techniques cannot be directly applied to graph data, as they often require fixed input data, while graph data often have varying sizes. Moreover, all these methods do not provide guaranteed watermarking performance under the attack. 

{\bf Backdoor-based watermarking for centralized GL models on graph data.}  
The only work \citep{xu2023watermarking} handling graph data uses random graph as the watermark.  
This method can be extended to the FedGL model, but its watermarking performance is far from satisfactory--especially vulnerable to watermark removal attacks such as distillation, finetuning, and the layer perturbation attack. 

{\bf Backdoor attacks for centralized and federated GL models on graph data.} Several works~\citep{zhang2021backdoor,yang2024distributed} design backdoor attacks to manipulate GL models, enabling the attacker to influence the learned model to serve its purpose--the  model will predict the attacker-chosen label for test graphs once they contain a predefined trigger. 
For instance, \cite{zhang2021backdoor} uses random subgraphs as
the trigger to backdoor centralized GL models, while 
\cite{yang2024distributed} learns subgraph trigger to  backdoor FedGL models. Note that the goal of these works is orthogonal to ours. 
\vspace{-4mm}
\section{Conclusion}
\vspace{-2mm}
We protect 
the model ownership of  emerging FedGL trained on distributed graph data, and use the \emph{de facto} backdoor-based watermarking method. 
We develop the first certifiably robust backdoor-based watermarking method {\name} for FedGL. 
{\name} demonstrates the capability of achieving high empirical watermarking performance under no attack, under existing backdoor removal attacks and the proposed stronger layer-perturbation  attack.   
{\name} is also provably robust against the worst-case layer-perturbation attack, once the number of perturbed layers is bounded by Theorem~\ref{thm:certify}.

{\bf Acknowledgments.} 
We thank all anonymous reviewers for
the constructive comments. Li is
partially supported by the National Natural Science Foundation of China under Grant No. 62072208, Key Research and Development Projects of Jilin Province under Grant No. 20240302090GX. 
Hong is partially supported by the National Science
Foundation under grant No. CNS-2302689, CNS-2308730, CNS-2319277 and CMMI-2326341.
Wang is partially supported by the National Science Foundation under grant No. ECCS-2216926, CNS-2241713, CNS-2331302 and CNS-2339686. 



%
%
\bibliographystyle{named}
\bibliography{egbib}
\appendix
\clearpage
\section{Proofs}
\label{app:thmproofs}
\noindent {\bf Proof of Theorem~\ref{thm:certified}:}  
Given a watermarked graph $G_w$, a base graph classifier $f$, our ensemble classifier $g$, and the target watermarked FedGL model $\theta = \{\vartheta_{i}\}_{i=1}^S$. 
Let $\theta'$ be the watermarked FedGL model after the layer-perturbation attack. 
We denote by $y=f(\vartheta_i,G_w)$ the prediction of $G_w$ by submodel $i$, and $g(\theta, G_w) = {\arg\max}_{y \in \mathbb{Y}} \ N_y$ the prediction of $G_w$ by the target model $\theta$, where $N_y={\textstyle \sum_{i=1}^S} \mathbbm{I}(f(\vartheta_{i},G_w) = y)$ and $\theta = \{\vartheta_{i}\}_{i=1}^S$. Assume $N_A$ and $N_B$ are the largest and second largest votes of the output of the ensemble classifier $g$ on the target model $\theta$ and $G_w$, respectively. Similarly, let $N_A'$ and $N_B'$ respectively denote the corresponding votes of the model $\theta'$. 
Under our GL model architecture, 
each perturbed layer affects at most $1$ submodel in the worst case. Therefore, $N_A'$ and $N_B'$ satisfy 
the following equations when $r$ layers in $\theta$ are perturbed: 
\begin{equation}
\begin{aligned}
    & N_A-r \le N'_A \le N_A+r, \\
    & N_B-r \le N'_B \le N_B+r. 
\label{equation: N_A - {N}_A & N_B - {N}_B}
\end{aligned}
\end{equation}
Thus, $g(\theta',G) = g(\theta,G)$ when $N_A' > N_B'$, This means $N_A-r > N_B+r - \mathbbm{1}(A<B)$, where $\mathbbm{I}(A<B)$ indicates we select a label with a smaller index in case of ties. 
Hence, $r$ satisfies:
\begin{equation}
r \le  r^* = 
\frac{N_A - N_B + \mathbbm{I}(A<B) -1}{2}. 
\end{equation}

\noindent {\bf Proof of Theorem~\ref{thm:tight}:} 
We use proof by contradiction, i.e., when $r > r^*$, we can build a base classifier $f'$ such that there exists $g(\theta,G_w) \neq  g(\theta',G_w)$. We select the smallest $r = r^* + 1$ for simplicity, indicating that $r^* + 1$ layers whose model parameters can be arbitrarily perturbed. 
Assume there exist a graph classifier $f'$ under which, $(r^*+1)$ submodels whose original predictions on $G_w$ are the label $A$ and now become $B$ after $r^*+1$ layers are perturbed.  
Then we have $N_A' = N_A - (r^* + 1)$ and $N_B' = N_B + (r^* + 1)$. Since $r^*$ is the largest number such that Equation (\ref{equation: N_A - {N}_A & N_B - {N}_B}) is satisfied, 
i.e., $N_A -r^* >   N_B - \mathbbm{1}(A<B) + r^*$ holds. Thus, 
\begin{equation}
N_A -(r^*+1) \leq   N_B - \mathbbm{1}(A<B) + (r^*+1). 
\end{equation}
Or $N_A' \leq N_B'$. Hence, $g(\theta,G_w) = A \ne B = g(\theta',G_w)$, if $A>B$. 

\section{Algorithm \ref{algorithm: training}}

\begin{algorithm}[H]
\caption{The training process of {\name}} 
\label{algorithm: training}
\small
\begin{flushleft}
{\bf Input:} clients $[1,T]$, watermarked clients $[1,T_w]$, training graphs $\mathbb{G}^i$ ($i \in [1,T]$), initial global model $\theta_1$, initial CWG $\omega_1$, \#Perturbed Layers $r$, 
unique key 
$k^i$ for client $i$ ($i \in [1,T_w]$), the number of submodels $S$. 

{\bf Output:} Target global model $\theta_{epoch}$, and target local models $\theta_{epoch}^j (j \in [1, T_w])$. 
\end{flushleft}
\begin{algorithmic}[1]
\For{each epoch $e$ in $[1,epoch]$}
    \For{each client $i \in [1,T]$} 
        \If{$i \in [1, T_w]$}
            \State Divide $\mathbb{G}^i$ into  $\mathbb{G}^i_c$ and $\mathbb{G}^i_{w}$.
            $\rhd$ clean \& watermarked samples
            \State $\mathbb{G}^i_{w}=\mathrm{CWG}(\mathbb{G}^i_{w},k^i)$
        \EndIf    
        \State $\vartheta_1, \vartheta_2, \cdots, \vartheta_S = \mathrm{divide}({\theta}_e)$ 
        \For{each submodel $j$ in $[1,S]$}
            \State $\vartheta^i_j = \underset{\vartheta^i_j}{\arg\min} \, L(\vartheta_j;\mathbb{G}^i)$ 
        \EndFor
        \State ${\theta}^{i}_e = \mathrm{joint}(\vartheta^i_1, \vartheta^i_2, \cdots, \vartheta^i_S)$
        \If{$i \in [1, T_w]$}
            \State $\omega^{i}_e =  \underset{\omega^{i}}{\arg\min} \, L(\theta^{i}_e; \mathbb{G}^i_{w})$
        \EndIf
    \EndFor
    \State $\theta_{e+1} =\frac{1}{|\mathbb{T}_e|}  {\textstyle \sum_{i \in \mathbb{T}_e}} \theta^{i}_e$
    $\rhd$ Server 
    selects $\mathbb{T}_e$ clients for aggregation
\EndFor
\end{algorithmic}
\end{algorithm} 

\section{More Experimental Details}
\label{app:experiment setup}

Table \ref{table:datasets} shows the statistics of the used graph datasets. Table \ref{app:CWG} shows the network architectures of the CWG component. 
Table \ref{app:model} shows the detailed network architectures of Fed-GIN, Fed-GSAGE, and Fed-GCN
models. 

\begin{table*}[!t]
\renewcommand{\arraystretch}{1.0}
\addtolength{\tabcolsep}{-2pt}
\caption{Statistics of datasets.}
\centering
    \begin{tabular}{c||c|c|c|c|c c c|c c c}
    \toprule
    \multicolumn{1}{c}{Datasets} & \multicolumn{1}{c}{\#Graphs} & \multicolumn{1}{c}{\#Classes} & \multicolumn{1}{c}{Avg. \#$\text{Node}$} & \multicolumn{1}{c}{Avg. \#$\text{
    Edge}$} & \multicolumn{3}{c}{\#Training graphs} & \multicolumn{3}{c}{\#Testing graphs} \\
    & & & & & 1 &   2 &   3 &   1 &   2 &   3 \\
    \hline
    \hline
    MUTAG & 188 & 2 & 17.93 & 19.80  & 83 & 42 & - &  42 & 21 & - \\
    PROTEINS & 1110 & 2 & 37.72 & 70.35  & 440 & 300 & -  & 220 &  150 &  - \\
    DD & 950 & 2 & 208.3 & 518.76 & 330 & 303 & -&   165 & 152  &  -  \\
    COLLAB & 4981 & 3 & 73.49 & 2336.66 & 517 & 1589 & 1215 & 258 & 794 & 608  \\
    \bottomrule
    \end{tabular}
\label{table:datasets}
\end{table*}

\begin{table*}[!t]
\renewcommand{\arraystretch}{1.0}
\addtolength{\tabcolsep}{0.1pt}
\caption{Detailed network architectures of CWG. }
\centering
    \begin{tabular}{c||c c c}
    \toprule
    Name & Network Architectures & Input & Output \\
    \hline
    \hline
    $\mathrm{GatingNet}$ & 2$\times$(Linear-$|\mathbb{V}|$\& ReLU \& Dropout-$0.05$) & ${\bf A}^i \in \{0,1\}^{|\mathbb{V}|\times |\mathbb{V}|}$ & ${\tilde{\bf A}}^i \in [0,1]^{|\mathbb{V}|\times |\mathbb{V}|}$\\
    & Linear-$|\mathbb{V}|$ \& Sigmoid & & \\
    $\mathrm{KeyNet}$ & 2$\times$(Linear-$|\mathbb{V}|$\& ReLU \& Dropout-$0.05$) & $ {\bf K}^i \in \mathbb{R}^{|\mathbb{V}|\times |\mathbb{V}|}$ & ${\tilde{\bf K}}^i \in [0,1]^{|\mathbb{V}|\times |\mathbb{V}|}$\\
    & Linear-$|\mathbb{V}|$ \& Sigmoid & & \\
    \bottomrule
    \end{tabular}
\label{app:CWG}
\end{table*}

\begin{table*}[!t]
\renewcommand{\arraystretch}{1.0}
\addtolength{\tabcolsep}{-5pt}
\caption{Detailed network architectures for $S=\{4,8,16\}$ in Fed-GIN, Fed-GSAGE, and Fed-GCN models within our watermark. The convolutional (conv) layers in these three models are GINConv, GSAGEConv, and GCNConv, respectively. Here, $x$ represents the input dimension, while $y$ represents the output dimension. BN is short for BatchNorm.} 
\footnotesize 
\centering
    \begin{tabular}{c|c|c c}
    \toprule
    submodels $1$-$4$  &  {submodels $5$-$8$} & \multicolumn{2}{c}{submodels $9$-$16$}  \\ 
    \hline
    \hline
    Linear-$x \times y$  & Linear-$x \times y$  & Linear-$x \times y$ & Linear-$x \times y$ \\
    conv-$x \times$64 \& BN-64 & conv-$x \times$128 \& BN-128 & conv-$x \times$72 \& BN-72 & conv-$x \times$96 \& BN-96  \\
    ReLU \& Linear-64$\times y$ & ReLU \& Linear-128$\times y$ & ReLU \& Linear-72$\times y$ & ReLU \& Linear-96$\times y$  \\

    2$\times$(conv-64$\times$64 \& BN-64 & 2$\times$(conv-128$\times$128 \& BN-128 & 2$\times$(conv-72$\times$72 \& BN-72 & 2$\times$(conv-96$\times$96 \& BN-96  \\
    ReLU \& Linear-64$\times y$) & ReLU \& Linear-128$\times y$) & ReLU \& Linear-72$\times y$) & ReLU \& Linear-96$\times y$)  \\
    
    conv-64$\times$64 \& BN-64 & conv-128$\times$64 \& BN-64 & conv-72$\times$64 \& BN-64 & conv-96$\times$64 \& BN-64 \\
    ReLU \& Linear-64$\times y$ & ReLU \& Linear-64$\times y$ & ReLU \& Linear-64$\times y$ & ReLU \& Linear-64$\times y$\\
    \hline
    Linear-$x \times y$  & Linear-$x \times y$  & Linear-$x \times y$ & Linear-$x \times y$ \\
    conv-$x \times$64 \& BN-64 & conv-$x \times$128 \& BN-128 & conv-$x \times$72 \& BN-72 & conv-$x \times$96 \& BN-96\\
    ReLU \& Linear-64$\times y$ & ReLU \& Linear-128$\times y$& ReLU \& Linear-72$\times y$& ReLU \& Linear-96$\times y$ \\

    2$\times$(conv-64$\times$64 \& BN-64 & 2$\times$(conv-128$\times$128 \& BN-128 & 2$\times$(conv-72$\times$72 \& BN-72 & 2$\times$(conv-96$\times$96 \& BN-96  \\
    ReLU \& Linear-64$\times y$) & ReLU \& Linear-128$\times y$) & ReLU \& Linear-72$\times y$) & ReLU \& Linear-96$\times y$)  \\
    
    conv-64$\times$32 \& BN-32& conv-128$\times$32 \& BN-32& conv-72$\times$32 \& BN-32& conv-96$\times$32 \& BN-32 \\
    ReLU \& Linear-32$\times y$ & ReLU \& Linear-32$\times y$ & ReLU \& Linear-32$\times y$ & ReLU \& Linear-32$\times y$\\
    \hline
    Linear-$x \times y$  & Linear-$x \times y$  & Linear-$x \times y$ & Linear-$x \times y$ \\
    conv-$x \times$64 \& BN-64 & conv-$x \times$128 \& BN-128 & conv-$x \times$72 \& BN-72 & conv-$x \times$96 \& BN-96\\
    ReLU \& Linear-64$\times y$ & ReLU \& Linear-128$\times y$ & ReLU \& Linear-72$\times y$ & ReLU \& Linear-96$\times y$\\
    2$\times$(conv-64$\times$64 \& BN-64 & 2$\times$(conv-128$\times$128 \& BN-128 & 2$\times$(conv-72$\times$72 \& BN-72 & 2$\times$(conv-96$\times$96 \& BN-96  \\
    ReLU \& Linear-64$\times y$) & ReLU \& Linear-128$\times y$) & ReLU \& Linear-72$\times y$) & ReLU \& Linear-96$\times y$)  \\
    conv-64$\times$16 \& BN-16 & conv-128$\times$16 \& BN-16 & conv-72$\times$16 \& BN-16 & conv-96$\times$16 \& BN-16\\
    ReLU \& Linear-16$\times y$ & ReLU \& Linear-16$\times y$ & ReLU \& Linear-16$\times y$ & ReLU \& Linear-16$\times y$\\
    \hline
    Linear-$x \times y$  & Linear-$x \times y$  & Linear-$x \times y$ & Linear-$x \times y$ \\
    conv-$x \times$64 \& BN-64 & conv-$x \times$128 \& BN-128 & conv-$x \times$72 \& BN-72 & conv-$x \times$96 \& BN-96 \\
    ReLU \& Linear-64$\times y$ & ReLU \& Linear-128$\times y$ & ReLU \& Linear-72$\times y$ & ReLU \& Linear-96$\times y$ \\
    2$\times$(conv-64$\times$64 \& BN-64 & 2$\times$(conv-128$\times$128 \& BN-128 & 2$\times$(conv-72$\times$72 \& BN-72 & 2$\times$(conv-96$\times$96 \& BN-96  \\
    ReLU \& Linear-64$\times y$) & ReLU \& Linear-128$\times y$) & ReLU \& Linear-72$\times y$) & ReLU \& Linear-96$\times y$)  \\
    conv-64$\times$8 \& BN-8 & conv-128$\times$8 \& BN-8 & conv-72$\times$8 \& BN-8 & conv-96$\times$8 \& BN-8 \\
    ReLU \& Linear-8$\times y$ & ReLU \& Linear-8$\times y$ & ReLU \& Linear-8$\times y$ & ReLU \& Linear-8$\times y$ \\
    \bottomrule
    \end{tabular}
\label{app:model}
\end{table*}

\section{More Experimental Results}

\begin{figure*}[!t]
\centering	
\includegraphics[width=\textwidth]{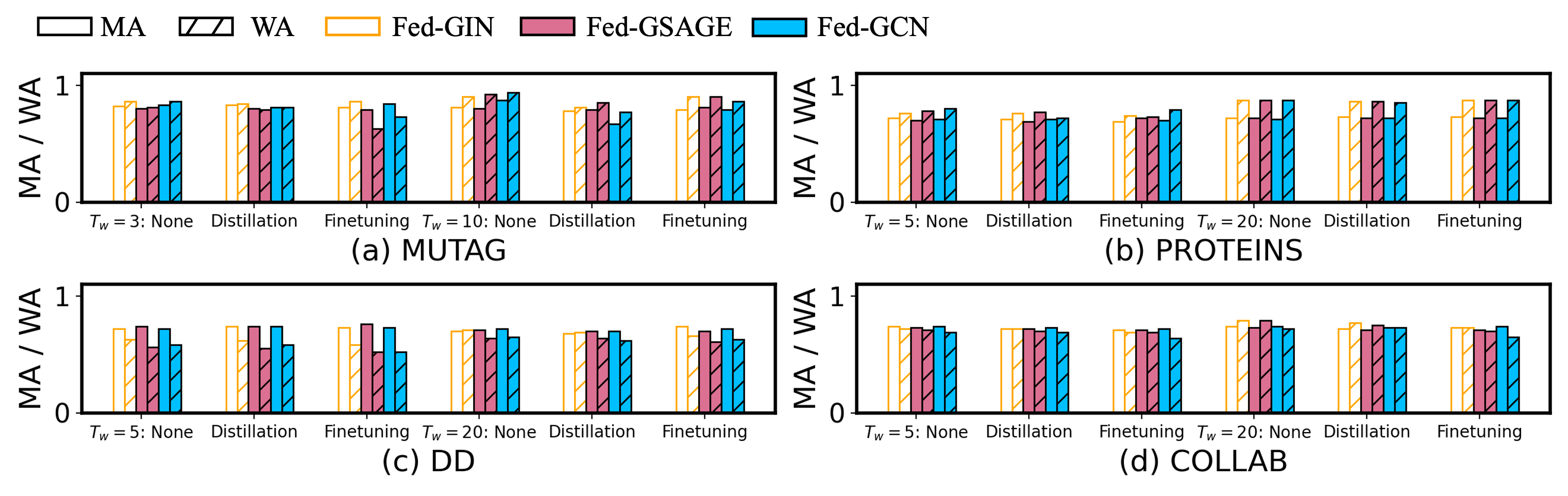}
    \vspace{-6mm}
	\caption{Impact of $T_w$ on FedGMark against prior watermark removal  and  layer-perturbation attacks.
 } 
        \label{fig:experiment-total-nc=5-and-20}
        \vspace{-2mm}
\end{figure*}

\begin{figure*}[!t]
\centering	
\includegraphics[width=\textwidth]{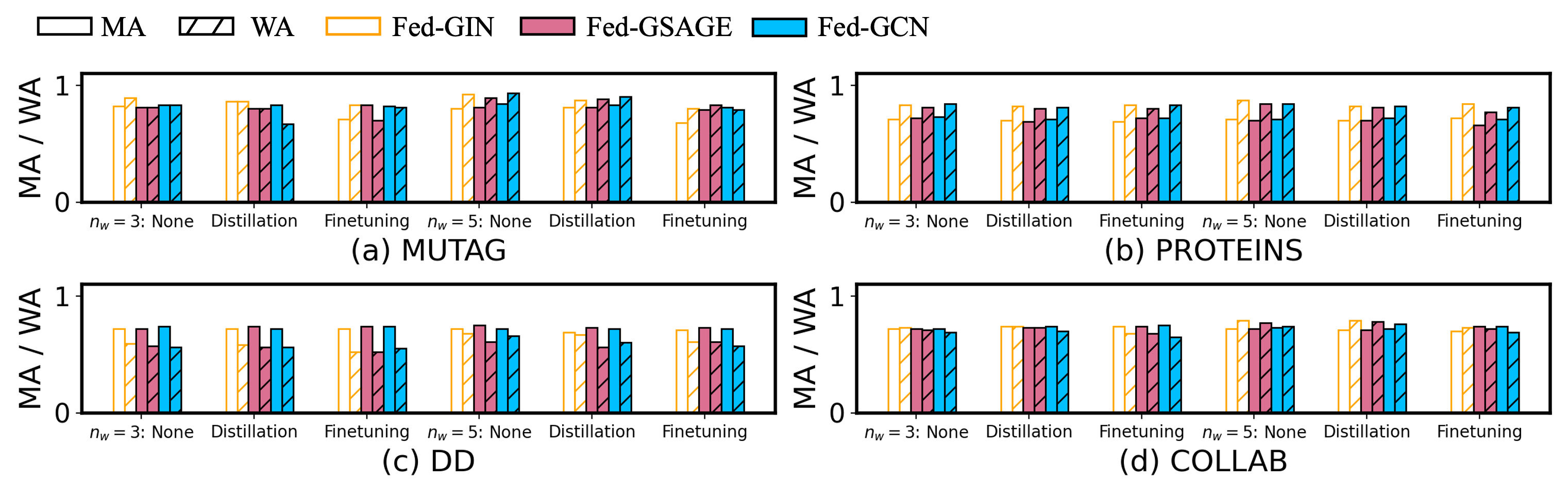}
   \vspace{-6mm}
	\caption{Impact of $n_w$ on FedGMark against prior watermark removal and  layer-perturbation attacks.
 } 
        \label{fig:experiment-total-ntri=3-and-5}
        \vspace{-2mm}
\end{figure*}

\begin{table}[!t]
\small 
\renewcommand{\arraystretch}{0.9}
\addtolength{\tabcolsep}{-3pt}
\caption{{Effect of learning rate ($lr$) and \#local epochs ($le$).}
}
\centering
    \begin{tabular}{c||c c c|c c c}
    \toprule
     & \multicolumn{3}{c}{$le=5$} & \multicolumn{3}{c}{$lr=0.01$}\\
     {Datasets} & $lr=0.01$ & $lr=0.05$ & $lr=0.1$ & $le=5$ & $le=10$ & $le=20$ \\
     & MA$\uparrow$ \ WA$\uparrow$ & MA$\uparrow$ \ WA$\uparrow$ & MA$\uparrow$ \ WA$\uparrow$ & MA$\uparrow$ \ WA$\uparrow$ & MA$\uparrow$ \ WA$\uparrow$ &  MA$\uparrow$ \ WA$\uparrow$  \\
    \hline
    \hline
    MUTAG &  0.81 \ \ 0.90  &  0.84 \ \ 0.89  &  0.79 \ \ 0.90  & 0.81 \ \ 0.90  &  0.84 \ \ 0.90  & 0.76 \ \ 0.92  \\
    PROTEINS & 0.72 \ \ 0.86   &  0.71 \ \ 0.79  & 0.71 \ \ 0.73   & 0.72 \ \ 0.86  &  0.72 \ \ 0.87  & 0.72 \ \ 0.89  \\
    DD &  0.73 \ \ 0.65  & 0.71 \ \ 0.57   &  0.70 \ \ 0.53  & 0.73 \ \ 0.65  & 0.73 \ \ 0.66   & 0.73 \ \ 0.69  \\
    COLLAB &  0.73 \ \ 0.75  & 0.72 \ \ 0.71   &  0.70 \ \ 0.65  & 0.73 \ \ 0.75  & 0.74 \ \ 0.75   & 0.73 \ \ 0.77  \\

    \bottomrule
    \end{tabular}
\label{app:table: le and lr}  
\end{table}

{\bf Impact of \#watermarking clients $T_w$ and watermark size $n_w$.} See results in Figure~\ref{fig:experiment-total-nc=5-and-20} and Figure~\ref{fig:experiment-total-ntri=3-and-5}, respectively. We can see {\name} achieves better watermarking performance with a larger number of clients generating watermarks; and a larger watermark size. 

{\bf {Impact of learning rate ($lr$) and \#local epochs ($le$).}} The results with varying $lr$ and $le$ are shown in Table~\ref{app:table: le and lr} in Appendix. We can see that a large $lr$ may reduce WA, and WA increases slightly as $le$ grows, indicating more thorough training makes our method perform better.

{\bf Scalability of FedGMark.} Compared with graph-based or non-robust watermarking methods, the computation overhead of FedGMark is mainly from the introduced submodel models (fixing all the other parameters, such as \#clients, \#iterations, to be the same). Particularly, the overhead scales linearly with the number of submodels $S$, and the runtime results are shown in Table~\ref{table:The Cost of RML}.

\begin{table}[!h]
\footnotesize
\renewcommand{\arraystretch}{1.0}
\addtolength{\tabcolsep}{-1.5pt}
\caption{Number of submodels $S$ vs. the runtime of RML. We observe an almost linear relationship between the total training time $C_t$ and the number of submodels $S$, expressed as $C_t \approx S*C_s$, where $C_s$ denotes the training time for an individual submodel.
}
\centering
    \begin{tabular}{c||c|c c c|c||c|c c c}
    \toprule
    {Datasets}  & $S$  &  4 & 8 & 16 &{Datasets}  & $S$ &    4 & 8 & 16\\ 
    & (Net) &  \multicolumn{3}{c|}{time(s)} & & (Net) & \multicolumn{3}{c}{time(s)}  \\
    \hline
    & Fed-GIN &   0.46  &  0.74 & 1.32  & & Fed-GIN &  2.79 & 5.10 & 9.49 \\
    MUTAG &  Fed-GSAGE &   0.28 &  0.48 &  0.90 & PROTEINS & Fed-GSAGE & 2.72 & 4.25 & 8.31    \\
    & Fed-GCN &  0.30  & 0.53 & 0.94  & & Fed-GCN & 2.73& 4.56& 8.95 \\
    \hline
    & Fed-GIN &  11.14  & 20.12 & 36.74  & & Fed-GIN  & 161.46 & 296.68  & 563.51 \\
    DD &  Fed-GSAGE &  11.75  & 19.82  & 31.57  & COLLAB & Fed-GSAGE & 152.23 & 281.78 & 551.86   \\
    & Fed-GCN &    11.61 & 19.51 & 33.23  & & Fed-GCN &  163.03 & 293.55 & 544.17 \\
    \bottomrule
    \end{tabular}
\label{table:The Cost of RML}  
\end{table}

{\bf {\name} under watermark/backdoor detection attacks.} 
Many existing works \citep{zhang2021backdoor} show the trigger-reverse based backdoor detection is ineffective to ``stealthy” backdoor. This is because the effectiveness of trigger reverse attacks largely depends on the statistical differences between clean data and backdoored data. Since we do not notice any graph backdoor trigger-reverse attack, we instead propose to quantitatively test the structure similarity between the generated watermarked graphs and the clean graphs. 
Here we use the metrics NetSim and DeltaCon proposed in \citep{wills2020metrics}, with the range $[0,1]$ and the higher value the larger similarity. As shown in Table~\ref{app:table: structure similarity}, we observe the watermarked graphs and their clean counterparts are structurally very close. This implies that the proposed watermarks are hard to be detected. 

{\bf {FedGMark on  non-IID/heterogeneous datasets.}} 
Recall that the local watermarks in {\name} are learnt by considering the unique properties in each client. Such unique properties may include the heterogeneity across clients’ data. 
To validate this, we also test {\name} with non-IID graphs across clients and show results in Table~\ref{app:table: NON-IID}, where each client holds a single label data. We observe that FedGMark demonstrates strong performance on non-IID datasets, indicating that the learnt customized watermarks effectively capture the heterogeneity of clients’ graphs.

\begin{table}[!t]
\small 
\renewcommand{\arraystretch}{1.1}
\addtolength{\tabcolsep}{12pt}
\caption{{Structure similarity of generated watermarked graphs and clean graphs. 
} }
\centering
    \begin{tabular}{c||c c c c}
    \toprule
    {Datasets} & MUTAG & PROTEINS & DD & COLLAB \\
    \hline
    \hline
    NetSim ($\uparrow$) & 0.97 & 0.98 & 0.99 & 0.99  \\
    DeltaCon ($\uparrow$) & 0.98 & 0.98 & 0.99 &0.99 \\
    \bottomrule
    \end{tabular}
\label{app:table: structure similarity}  
\end{table}

\begin{table}[!t]
\small 
\renewcommand{\arraystretch}{1.1}
\addtolength{\tabcolsep}{9pt}
\caption{{Results of FedGMark on IID and non-IID/heterogeneous   datasets. }
}
\centering
    \begin{tabular}{c||c c c c}
    \toprule
    {Datasets} & MUTAG & PROTEINS & DD & COLLAB \\
     & MA$\uparrow$ \ WA$\uparrow$ & MA$\uparrow$ \ WA$\uparrow$ & MA$\uparrow$ \ WA$\uparrow$ & MA$\uparrow$ \ WA$\uparrow$ \\
    \hline
    \hline
    IID & 0.81 \ \ 0.90  & 0.72 \ \ 0.86 &  0.73 \ \ 0.65 & 0.73 \ \ 0.75  \\
    Non-IID & 0.80 \ \ 0.89  & 0.72 \ \ 0.83  & 0.72 \ \ 0.63  & 0.72 \ \ 0.75  \\
    \bottomrule
    \end{tabular}
\label{app:table: NON-IID}  
\end{table}

\begin{table}[!t]
\small 
\centering
\renewcommand{\arraystretch}{1.1}
\addtolength{\tabcolsep}{9pt}
\caption{Results of FedGMark against $p\%$ malicious clients whose watermarked data are mislabeled.}
 \begin{tabular}{c||c c c c c}
 \toprule
   & 0\%      & 10\%     & 20\%     & 30\%     & 40\%     \\ 
MUTAG & .81/.90  & .80/.90  & .80/.89  & .80/.80  & .80/.71  \\ 
PROTEIN & .72/.86  & .72/.85  & .72/.76  & .71/.67  & .70/.67  \\ 
DD & .73/.65  & .72/.63  & .72/.58  & .71/.53  & .71/.52  \\ 
COLLAB & .73/.75  & .74/.74  & .73/.68  & .73/.61  & .72/.60  \\ 
 \bottomrule
\end{tabular}
\label{tbl:strongerattack}
\vspace{-4mm}
\end{table}

{\bf {FedGMark against more knowledgeable/stronger adversaries.}} 
In our threat model, we assume all clients and the server are benign and assume the attacker does not know our CWG. Here, we test FedGMark against stronger adversaries, where some clients are malicious and the these malicious clients also has access to the CWG component to manipulate the FedGL training. 
First, we consider a \emph{passive} attack where all malicious clients \emph{do not} use CWG to generate customized local watermarks. Model (b) in Table \ref{table:ablation study} shows the \emph{maximum} WA decrease is 9\%, where \emph{all} clients do not use CWG.   

Second, we test an \emph{active} attack where malicious clients modify their watermark data's label to obfuscate the training. Specifically, all malicious clients’ watermark data are labeled (e.g., 2) differently from the target label (e.g., 1) and then follow the federated training. The results in Table \ref{tbl:strongerattack} show MA/WA is marginally affected even with 20\% malicious clients.

{\bf {FedGMark with alternative aggregation methods.}} 
The current FedGMark's evaluation focuses on FedAvg for aggregating client models. Here, we evaluate FedGMark using Multi-Krum~\citep{blanchard2017machine} and Trim-mean~\citep{yin2018byzantine} aggregation methods that consider data quality (e.g., remove outlier clients). Specifically, Multi-Krum filters a set of $p$ clients whose gradients largely deviated from others, while Trim-mean trims off $q$ highest and lowest values for each parameter in clients' models.
Table~\ref{app:table: Other aggregation methods} shows the results with $p=10$ and $q=10$. 
We can see these robust aggregators achieve a robustness-utility tradeoff, and MA and WA are not largely different.

\begin{table}[!t]
\small 
\renewcommand{\arraystretch}{1.1}
\addtolength{\tabcolsep}{9pt}
\caption{{Results of our FedGMark against Multi-Krum~\citep{blanchard2017machine} and Trim-mean~\citep{yin2018byzantine} aggregation methods. }
}
\centering
    \begin{tabular}{c||c c c c}
    \toprule
    {Datasets} & MUTAG & PROTEINS & DD & COLLAB \\
     & MA$\uparrow$ \ WA$\uparrow$ & MA$\uparrow$ \ WA$\uparrow$ & MA$\uparrow$ \ WA$\uparrow$ & MA$\uparrow$ \ WA$\uparrow$ \\
    \hline
    \hline
    Avg &  0.81 \ \ 0.90  & 0.72 \ \ 0.86  &  0.73 \ \ 0.65  & 0.73 \ \ 0.75   \\
    Multi-Krum &  0.78 \ \ 0.92  & 0.73 \ \ 0.85  &  0.72 \ \ 0.63  & 0.73 \ \ 0.74   \\
    Trim-mean &  0.75 \ \ 0.93  & 0.70 \ \ 0.87  &  0.70 \ \ 0.65  &  0.71 \ \ 0.77  \\
    \bottomrule
    \end{tabular}
\label{app:table: Other aggregation methods}  
\end{table}

{\bf {FedGMark with alternative triggers such as feature-based triggers and hybrid triggers.}} 
The current FedGMark relies on graph structural information to learn the trigger used in watermark. Here, we adjust FedGMark to learn feature-based triggers and hybrid feature-structure triggers.

To learn feature-based triggers, we first select a set of nodes from a graph as the target nodes, and learn the watermarked features for the target nodes (we do not watermark structure). We use a graph $G^i = (\mathbb{V}^i, \mathbb{E}^i, {\bf X}^i) $  from client $i$ for illustration, where ${\bf X}^i$ is the node feature matrix. We then define a feature-mask ${\bf M}_f^i[v_j]=1$ if $v_j \in \mathbb{V}_w^i$ and 0 otherwise, where $\mathbb{V}_w^i$ is the watermark node set described in the paper. Then, we introduce a feature network ($\mathrm{FeaNet}$) that learns watermarked node features as $\mathbf{X}_w^i = \mathrm{FeaNet}({\bf X}^i) \odot {\bf M}_f^i$. The $\mathrm{FeaNet}$ takes input ${\bf X}^i$ and outputs a matrix having the same size as ${\bf X}^i$, e.g., it has the same architecture as GatingNet but adjusts the input size. The corresponding watermarked graph is defined as  $G_w^i=(\mathbb{V}^i, \mathbb{E}^i, \mathbf{X}_w^i)$. By generating a set of watermarked graphs {$G_w^i$} for client $i$, we minimize the loss on client $i$’s both clean graphs {$ G_c^i$} and {$G_w^i$}. 

Further, to learn feature-structure triggers, we combine $\mathrm{FeaNet}$ (that gets $\mathbf{X}_w^i$) with $\mathrm{GatingNet}$/$\mathrm{KeyNet}$ (that gets $\mathbb{E}_w^i$), and the watermarked graphs are $G_w^i=(\mathbb{V}^i, \mathbb{E}_w^i, \mathbf{X}_w^i)$. We then minimize the loss on {$G_c^i$} and {$G_w^i$}. More details about training refer to Section 3.4.     

We evaluate these triggers and the results are in Table~\ref{tbl:altertriggers}. We observe that structure information alone is sufficient to enable designing effective triggers.  

\begin{table}[!t]
\small 
\centering
\renewcommand{\arraystretch}{1.1}
\addtolength{\tabcolsep}{9pt}
\caption{{\name} with alternative triggers.}
    \begin{tabular}{c||c c c c}
    \toprule
{\bf Datasets} & \textbf{feature} & \textbf{structure} & \textbf{feature-structure} \\ \hline \hline
MUTAG & .81/.78 & .81/.90 & .79/.92 \\ 
PROTEIN & .72/.77 & .72/.86 & .73/.87 \\ 
DD & .72/.53 & .73/.65 & .74/.66 \\ 
COLLAB & .73/.67 & .73/.75 & .72/.76 \\ 
\bottomrule
\end{tabular}
\label{tbl:altertriggers}
\end{table}

\begin{figure*}[!t]
\centering	
\includegraphics[width=\textwidth]{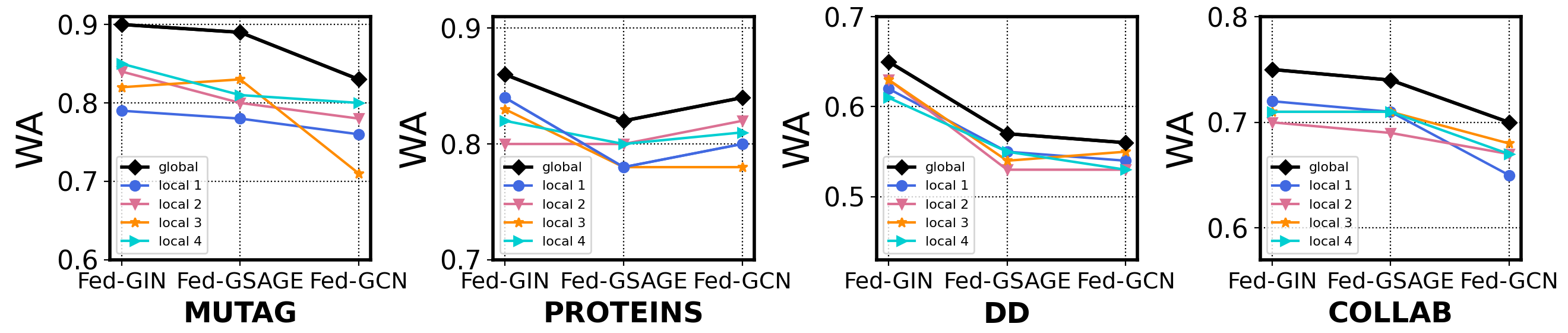}
    \vspace{-6mm}
	\caption{Global watermark vs. local watermarks $1$-$4$.} 
        \label{app:fig:local trigger}
        \vspace{-4mm}
\end{figure*}

\section{Discussion}
\vspace{-1mm}

{\bf Global watermark vs. local watermarks.} 
In this paper, each client employs CWG to generate watermarks (i.e., local watermarks) to train the local model. During testing, we can use either local watermarks or a global watermark, i.e., fully connect local watermarks. 
Here, we further evaluate the impact of global watermark vs. local watermarks of the target watermarked  model under default settings. For better clarity, we randomly select four local watermarks labeled as local watermarks $1$ to $4$. The experimental results depicted in Figure~\ref{app:fig:local trigger} demonstrate that, despite not being part of the training process, the global watermark performs slightly better than local watermarks. This implies that the distributed training of FedGL can aggregate the effects of local watermarks, providing a foundation for utilizing the global watermark for verification.

{\bf Ensemble classifier with majority voting vs. summation on submodels' outputs.} 
In this paper,  
we propose an ensemble classifier $g$ to aggregate the predictions of each submodel $\vartheta_{i}$ through a voting mechanism and classify a graph $G$ as  $g(\theta,G)={\arg\max}_{y \in \mathbb{Y}} N_y$, where
$ N_y = {\textstyle \sum_{i=1}^S} \mathbbm{I}(f(\vartheta_{i},G) = y)$. Another common strategy, as used in GL models such as GIN~\citep{xu2018powerful}, is to 
sum the prediction \emph{confidence vector} of each submodel $\vartheta_{i}$, i.e., $g(\theta,G)=
{\arg\max}_{y \in \mathbb{Y}} {\bf f}_G$, where ${\bf f}_G
= {\textstyle \sum_{i=1}^S}{\bf f}(\vartheta_{i},G)$ and 
${\bf f}$ outputs a probabilistic confidence vector whose summation is 1. 
Here, we also test our {\name} with this strategy against the 1-layer perturbation attack, and the results are shown in Table \ref{app:table:normal classifier}. 
We notice the results  
do not remain consistent across different numbers of submodels and are worse than those in Table~\ref{table:experiment-total-8-and-16}, especially when $S$ is small. 
This suggests that the summing-based ensemble classifier is not robust to the layer-perturbation attack, emphasizing the necessity of our proposed majority-voting based ensemble classifier.

{\bf {\name} against offline clients that cannot provide watermark data or malicious clients that provide fake watermark samples during ownership verification.}
We deem that our ownership verification is still robust against offline clients and malicious clients, if its number is less than 50\%. During ownership verification, each client provides its own watermark data to the trusted judge. When some clients are offline, the trusted judge can simply neglect them and only use participating clients’ watermark data for verification. 

When facing malicious clients, their negative effect can be circumvented through a majority voting-based approach. Specifically, all clients provide their own watermark data to the judge and obtain the watermark accuracy per client. Though the watermark accuracy on malicious clients could be very low, the majority of benign clients can produce more number of high watermark accuracy, compared to the number of low accuracy. When the judge uses the majority-vote strategy, the final watermark accuracy is still high, ensuring the accurate ownership claim for benign clients.

{\bf Using clients' training graphs as the watermark samples and sending them to the trust judge may lead to privacy leakage.}
We clarify that the watermark data are not necessarily generated from the training/test samples. Remember the primary goal of backdoor-based watermarking is to force the model to memorize the relationship between the backdoor trigger (in the watermark samples) and the target label, while the samples to inject the trigger do not have constraints, i.e., they can be from training samples or artificially synthesized (which does not contain privacy information of any training/test data). For conveniences, existing methods inject backdoor triggers into the training/test samples. To validate this, we synthesize a set of random graphs using the popular Erdős–Rényi model (via the NetworkX toolbox) and the watermark samples are generated by injecting the learnt watermark on the synthesized graphs. Under the default setting, we test on Fed-GIN and show results in Table~\ref{tbl:syngraphs}, where we observe WAs are very close to those shown in the paper on the four datasets.

\begin{table}[!t]
\scriptsize 
\renewcommand{\arraystretch}{1.0}
\addtolength{\tabcolsep}{-0.7pt}
\caption{FedGMark with confidence scores summation against layer-perturbation attack. 
} 
\centering
    \begin{tabular}{c||c|c c c c||c|c c c}
    \toprule
    {Datasets}  & \multicolumn{1}{c}{$S$}&  {Fed-GIN}  &  {Fed-GSAGE} & {Fed-GCN} &{Datasets}  & \multicolumn{1}{c}{$S$}&  {Fed-GIN}  &  {Fed-GSAGE} & {Fed-GCN}  \\ 
    & & MA$\uparrow$ \ WA$\uparrow$ & MA$\uparrow$ \ WA$\uparrow$  & MA$\uparrow$ \ WA$\uparrow$  & & & MA$\uparrow$ \ WA$\uparrow$  & MA$\uparrow$ \ WA$\uparrow$  & MA$\uparrow$ \ WA$\uparrow$ \\
    \hline
    \hline
    & 4  & 0.71 \ \ 0.55 & 0.73 \ \ 0.65 & 0.81 \ \ 0.61 & &4  & 0.52 \ \ 0.51 & 0.53 \ \ 0.52 & 0.63 \ \ 0.67 \\
    MUTAG & 8  & 0.70\ \ 0.65  & 0.71\ \ 0.60 & 0.73\ \ 0.44 & PROTEINS& 8  & 0.70\ \ 0.83 & 0.71\ \ 0.84 & 0.73\ \ 0.86  \\
    & 16  &  0.81\ \ 0.83 &  0.81\ \ 0.76 & 0.79\ \ 0.75 &   & 16 &  0.72\ \ 0.85 &  0.72\ \ 0.85 &  0.71\ \ 0.83 \\
    \hline 
    & 4  &  0.71 \ \ 0.39  & 0.70 \ \ 0.28 & 0.73 \ \ 0.43 & & 4  & 0.69 \ \ 0.48  &0.67 \ \ 0.52  & 0.66 \ \ 0.51\\  
    DD& 8  & 0.70\ \ 0.42  & 0.72\ \ 0.48  & 0.73\ \ 0.46 & COLLAB &8  & 0.74 \ \ 0.61  & 0.74 \ \ 0.66  &  0.72 \ \ 0.62 \\
    & 16 &  0.73\ \ 0.59 &  0.74\ \ 0.54 & 0.71\ \ 0.42 &  &16  &  0.71 \ \ 0.75 &  0.71 \ \ 0.72 & 0.72 \ \ 0.70  \\
    \bottomrule
    \end{tabular}
\label{app:table:normal classifier}  
\end{table}

\begin{table}[!t]
\small
\centering
\renewcommand{\arraystretch}{1.0}
\addtolength{\tabcolsep}{-0.7pt}
\caption{MA/WA on synthesized graphs for watermarking.}
 \begin{tabular}{l||c c c c}
 \toprule
\textbf{Watermark} & \textbf{MUTAG} & \textbf{PROTEINS} & \textbf{DD} & \textbf{COLLAB} \\ \hline
on train/test graphs    & 0.81 / 0.90   & 0.72 / 0.86   & 0.73 / 0.65   & 0.73 / 0.75   \\ 
on synthesized graphs   & 0.80 / 0.88   & 0.71 / 0.84   & 0.72 / 0.64   & 0.72 / 0.73   \\ 
\bottomrule
\end{tabular}
\label{tbl:syngraphs}
\vspace{-2mm}
\end{table}

Furthermore, since all clients intend to verify model ownership, it is reasonable to believe that these clients are willing to provide their watermark data—whether generated from private training/test data or non-private synthesized data—exclusively to a trusted judge, with informed consent and in accordance with legal and ethical standards. From this perspective, the data is confidential between each client and the trusted judge. We acknowledge it is very interesting future work to design a provably private mechanism for model ownership verification that the verifier cannot access the watermark data but can guarantee the correctness of verification.

\clearpage
\newpage
\section*{NeurIPS Paper Checklist}

\begin{enumerate}

\item {\bf Claims}
    \item[] Question: Do the main claims made in the abstract and introduction accurately reflect the paper's contributions and scope?
    \item[] Answer: \answerYes{} 
    \item[] Justification: {We clearly state the research problem, motivation, and contributions of this paper in the Abstract and Introduction. }

\item {\bf Limitations}
    \item[] Question: Does the paper discuss the limitations of the work performed by the authors?
    \item[] Answer: \answerYes{} 
    \item[] Justification: {We discuss the limitations of the approach in the Discussion section, e.g., using clients' training graphs as watermark samples and sending them to the trusted judge could lead to privacy leakage.
    }

\item {\bf Theory Assumptions and Proofs}
    \item[] Question: For each theoretical result, does the paper provide the full set of assumptions and a complete (and correct) proof?
    \item[] Answer: \answerYes{} 
    \item[] Justification: {We provide complete theory background in the main text, and present proofs in the supplementary.}

    \item {\bf Experimental Result Reproducibility}
    \item[] Question: Does the paper fully disclose all the information needed to reproduce the main experimental results of the paper to the extent that it affects the main claims and/or conclusions of the paper (regardless of whether the code and data are provided or not)?
    \item[] Answer: \answerYes{} 
    \item[] Justification: {Our experimental results are reproducible. We provide detailed experimental settings in the main text.}

\item {\bf Open access to data and code}
    \item[] Question: Does the paper provide open access to the data and code, with sufficient instructions to faithfully reproduce the main experimental results, as described in supplemental material?
    \item[] Answer: \answerYes{} 
    \item[] Justification: {We include links to the source code in the abstract.}

\item {\bf Experimental Setting/Details}
    \item[] Question: Does the paper specify all the training and test details (e.g., data splits, hyperparameters, how they were chosen, type of optimizer, etc.) necessary to understand the results?
    \item[] Answer: \answerYes{} 
    \item[] Justification: {We provide training and test details in the experimental part.}

\item {\bf Experiment Statistical Significance}
    \item[] Question: Does the paper report error bars suitably and correctly defined or other appropriate information about the statistical significance of the experiments?
    \item[] Answer: \answerNo{} 
    \item[] Justification: {Our experimental results are stable. }

\item {\bf Experiments Compute Resources}
    \item[] Question: For each experiment, does the paper provide sufficient information on the computer resources (type of compute workers, memory, time of execution) needed to reproduce the experiments?
    \item[] Answer: \answerYes{} 
    \item[] Justification: {All experiments can be reproduced using one NVIDIA GeForce GTX 1080 Ti GPU.}
    
\item {\bf Code Of Ethics}
    \item[] Question: Does the research conducted in the paper conform, in every respect, with the NeurIPS Code of Ethics \url{https://neurips.cc/public/EthicsGuidelines}?
    \item[] Answer: \answerYes{} 
    \item[] Justification: {This research conforms to the NeurIPS Code of Ethics.}

\item {\bf Broader Impacts}
    \item[] Question: Does the paper discuss both potential positive societal impacts and negative societal impacts of the work performed?
    \item[] Answer: \answerYes{} 
    \item[] Justification: {Positive societal impacts: Our proposed watermark i.e. FedGMark can protect the ownership of FedGL models.}
    
\item {\bf Safeguards}
    \item[] Question: Does the paper describe safeguards that have been put in place for responsible release of data or models that have a high risk for misuse (e.g., pretrained language models, image generators, or scraped datasets)?
    \item[] Answer: \answerNA{} 
    \item[] Justification: {This paper poses no the above risks.}

\item {\bf Licenses for existing assets}
    \item[] Question: Are the creators or original owners of assets (e.g., code, data, models), used in the paper, properly credited and are the license and terms of use explicitly mentioned and properly respected?
    \item[] Answer: \answerYes{} 
    \item[] Justification: {We cite the reference of datasets employed in this paper following their requirements.}

\item {\bf New Assets}
    \item[] Question: Are new assets introduced in the paper well documented and is the documentation provided alongside the assets?
    \item[] Answer: \answerNA{} 
    \item[] Justification: {There is no dataset contribution in this paper.}

\item {\bf Crowdsourcing and Research with Human Subjects}
    \item[] Question: For crowdsourcing experiments and research with human subjects, does the paper include the full text of instructions given to participants and screenshots, if applicable, as well as details about compensation (if any)? 
    \item[] Answer: \answerNA{} 
    \item[] Justification: {Our work does not involve crowdsourcing and research with human subjects.}

\item {\bf Institutional Review Board (IRB) Approvals or Equivalent for Research with Human Subjects}
    \item[] Question: Does the paper describe potential risks incurred by study participants, whether such risks were disclosed to the subjects, and whether Institutional Review Board (IRB) approvals (or an equivalent approval/review based on the requirements of your country or institution) were obtained?
    \item[] Answer: \answerNA{} 
    \item[] Justification: {This paper does not involve crowdsourcing nor research with human subjects.}

\end{enumerate}

\end{document}